\documentclass[reprint,amsmath,amssymb,aps,prx,groupedaddress,superscriptaddress]{revtex4-1}
\usepackage{graphicx,xcolor,tikz}
\usepackage{amsthm,amssymb,amsmath,dsfont,braket}
\usepackage{bm}
\usepackage{hyperref} 
\usepackage{epstopdf,psfrag}
\usepackage{relsize,amsbsy}
\usepackage{multirow}
\DeclareMathOperator{\tr}{Tr}

\newcommand{\be}{\begin{equation}}
\newcommand{\ee}{\end{equation}}

\newcommand{\bit}{\begin{enumerate}}
\newcommand{\eit}{\end{enumerate}}

\definecolor{bananayellow}{rgb}{1.0, 0.88, 0.21}
\definecolor{straw}{rgb}{0.75, 0.71, 0.3}

\definecolor{palatinatepurple}{rgb}{0.49, 0.24, 0.46}

\definecolor{darkblue}{rgb}{0.0, 0.0, 0.55}

\definecolor{darkgreen}{rgb}{0.0, 0.5, 0.0}

\newcommand{\diff}{\mathrm{d}}

\usepackage[normalem]{ulem}

\newcommand{\ud}{\mathrm{d}}

\newcommand{\lefta}{\left\langle}
\newcommand{\righta}{\right\rangle}

\newcommand{\nn}{\nonumber}

\newcommand{\T}{{\mkern-1.5mu\mathsf{T}}}

\newcommand{\ie}{i.e.~}

\newcommand{\RNum}[1]{\uppercase\expandafter{\romannumeral #1\relax}}
\newcommand{\tx}[1]{\textmd{#1}}
\newcommand{\onefrac}[1]{\frac{1}{#1}}
\newcommand{\refeq}[1]{\textmd{Eq.\ }(\ref{#1})}
\newcommand{\reffg}[1]{\textmd{Fig.\ }\ref{#1}}
\newcommand{\refapp}[1]{\textmd{Appendix\ }\ref{#1}}

\definecolor{green}{HTML}{2ca02c}
\hypersetup{
    colorlinks,
    linkcolor={blue},
    citecolor={red},
    urlcolor={blue},
    linktocpage=true
}

\newcommand{\vrr}{\mathbf{r}}
\newcommand{\vq}{\mathbf{q}}

\newcommand{\vk}{\mathbf{k}}
\newcommand{\ve}[1]{{\bf #1}}

\newcommand{\Tr}{\operatorname{Tr}}

\begin{document}

\title{ Sound attenuation in the hyperhoneycomb  Kitaev spin liquid}

	\author{Kexin Feng}
	\affiliation{School of Physics and Astronomy, University of Minnesota, Minneapolis, Minnesota 55455, USA}	

		\author{Aysel Shiralieva}
\affiliation{Ecole normale superieure, Paris, France}

	\author{Natalia B. Perkins}
	\affiliation{School of Physics and Astronomy, University of Minnesota, Minneapolis, Minnesota 55455, USA}
	\date{\today}

	\begin{abstract} 
  In recent years,  it has been  shown that  the phonon dynamics may serve as an indirect probe of fractionalization of spin degrees of freedom.  Here we propose that the sound attenuation measurements allows for the characterization and identification of  the Kitaev quantum spin liquid on the hyperhoneycomb lattice, which is particularly interesting since the strong  Kitaev interaction was observed  in the the hyperhoneycomb  magnet $\beta$-Li$_2$IrO$_3$.  To this end we consider the low-temperature scattering between acoustic phonons  and gapless Majorana fermions  with nodal-line band structure. We  find  that the sound attenuation has  a characteristic angular dependence,  which is explicitly shown for the high-symmetry planes at temperatures below the flux energy gap.

	\end{abstract}	

	\maketitle

	\section{Introduction}

Quantum spin liquids (QSLs) are states of matter  in which no
symmetry is broken.  QSLs are interesting in general because they  exhibit
a remarkable set of collective phenomena  including
topological ground-state degeneracy, long-range entanglement, and fractionalized excitations \cite{Lee2008,Balents2010,Savary2016, Zhou2017,KnolleMoessner2019,Broholm2020,Motome2019}.
In recent years, much  work has  been done to
understand the nature of QSLs.  However,  this is not  generically an easy task since QSLs in realistic  models are usually ensured by
frustration, either from a particular  geometry of the lattice structure or from competing spin interactions, even identifying the  models which  host  such states is challenging.
In this sense, the exactly solvable Kitaev model on the honeycomb lattice with QSL ground state \cite{Kitaev2006}  and its possible realization in strongly spin-orbit couple materials \cite{Jackeli2009,Chaloupka2010} helped us  both with getting a   deeper insight  in the nature of QSL  state and  developing new approaches for detection of  this exotic phase of matter in experiment. 
A promising route for  searching for QSL physics in real materials is to look for signatures of spin fractionalizations in various types of  dynamical probes, such as inelastic neutron scattering,  Raman scattering,  resonant inelastic x-ray scattering, ultrafast spectroscopy   and terahertz non-linear coherent spectroscopy \cite{KnolleMoessner2019,Takagi2019,Broholm2020,Motome2019}. A possibility to compute  the corresponding response functions analytically  in the Kitaev model provides a unique opportunity to
 explore the characteristic fingerprints of the QSL physics in these dynamical probes on a more quantitative level \cite{Knolle2014a,Knolle2014b,Knolle2015,Brent2015,Brent2016-long,Smith2016,Gabor2016,Halasz2017,Gabor2019,Eschmann2020}.
 This is highly significant,  because  it gives us an  opportunity  to learn about generic behavior  of other QSLs, which are much more difficult  to describe. 

 It was recently shown that  a lot of  information can be  obtained   by studying the phonon dynamics in  the QSL candidate materials \cite{Plee2011,Plee2013,Metavitsiadis2020,Ye2020,Feng2021,Metavitsiadis2021}, since
the spin-lattice coupling is inevitable and often rather strong in real materials \cite{Hentrich2018,Kasahara2018,Pal2020,Miao2020}. 
The characteristic modifications of the  phonon dynamics from QSL compared with their non-magnetic or magnetically ordered analogs  can be probed in various observables, including
the renormalization of the spectrum of acoustic phonons \cite{Miao2020}, particular temperature dependence of the sound attenuation pattern and the phonon Hall viscosity\cite{Metavitsiadis2020,Ye2020,Feng2021}, the Fano lineshapes in the optical phonon Raman spectrum caused by the overlapping of the optical phonon  peaks with the continuum of the fractionalized excitations \cite{Sandilands2015,Glamazda2017,Mai2019,Dirk2020,Lin2020,Metavitsiadis2021,Kexin2021}, thermal conductivity and thermal Hall effect \cite{Kasahara2018}. Again, the presence of the exact solution of the  Kitaev  model helps to quantitatively understand the  dynamics of the phonons coupled to the underlying QSL. 

The Kitaev model 
can be generalized and defined for various three-coordinated  three-dimensional lattices \cite{Mandal2009, Hermanns2014,Hermanns2015,Hermanns2016,
Brent2015, 
Brent2016-long,Halasz2017,Trebst2017,Eschmann2020}, including the hyperhoneycomb, stripyhoneycomb, hyperhexagon, and
hyperoctagon lattices.  As a two-dimensional counterpart, these models   are exactly  solvable and have QSL ground state  with fractionalized excitations that are 
gapless Majorana fermions and gapped $Z_2$ gauge fluxes for the isotropic coupling parameters.
Importantly, the Majorana fermions  exhibit a rich variety of nodal structures due to
the different (projective) ways symmetries can act on them
\cite{Hermanns2014,Hermanns2015}.  These nodal structures include
nodal lines for the hyperhoneycomb and the
stripyhoneycomb models \cite{Mandal2009}, Fermi surfaces for the
hyperoctagon model \cite{Hermanns2014}, and the Weyl points for the
hyperhexagon model \cite{Hermanns2015}.

  In this work we performed a study of the phonon dynamics in the  Kitaev  model on the hyperhoneycomb lattice, which 
is particularly important among three-dimensional Kitaev models because  of  the existence  of the Kitaev candidate material $\beta$-Li$_2$IrO$_3$ \cite{Biffin2014a,Takayama2015,Ruiz2017,Ruiz2021,Yang2022,Halloran2022}, which is realized  on the hyperhoneycomb lattice.  While we know that  other interactions are present in this compound in addition to the dominant Kitaev interaction, here we assume that some  good intuition can be obtained by  studying the limiting case of the  pure Kitaev model. 
 To this end, we  derived  the Majorana fermion-phonon  coupling vertices using the symmetry considerations and used them for computation of the  phonon attenuation.

 The rest of the paper is organized as follows.	
In Sec.~\ref{ModelSec}, we  present the  derivation of the  spin-phonon Kitaev Hamiltonian on the hyperhoneycomb lattice.   We start by reviewing the  Kitaev  spin model on the hyperhoneycomb lattice
in Sec.~\ref{Kitaevmodel}. We obtain its fermionic band structure and  show that the fermions are gapless along the nodal line within the $\Gamma$-$X$-$Y$ plane, for which we obtain an analytical equation.
In Sec.~\ref{Sec:phonons}, we  introduce the lattice Hamiltonian for acoustic phonons on the hyperhoneycomb lattice.
 We obtain the acoustic phonon spectrum for the $D_{2h}$ point group symmetry   in the long wavelength approximation. 
In Sec.~\ref{Sec:MFPhcoupling},  we present the  explicit microscopic derivation of the Majorana fermion-phonon (MFPh) coupling vertices and show that there are four symmetry channels which contribute into them.
The knowledge of  the MFPh couplings allows us to  compute  the phonon dynamics, so
 we use the diagrammatic techniques and in  Sec.~\ref{sec:polarizationbubble} compute the phonon 
 polarization bubble. 
In Sec.~\ref{Sec:atten}, 
we  present our numerical results for  the attenuation coefficient  for acoustic phonon modes. To  this end, we  first discuss the  kinematic constraints in Sec.~\ref{Sec:kinematic} and then in Sec.~\ref{Sec:numerical}
 analyze the angular dependence of the sound
attenuation coefficient  for acoustic phonons with different polarizations.
In Sec.\ref{sec:discussion}, we present a  short  summary and discuss the possibility for  the spin fractionalization  in the Kitaev hyperhoneycomb model to be seen in the sound attenuation  measurements by the ultrasound experiments.

 \begin{figure}
		\centering
		\includegraphics[width=1\linewidth]{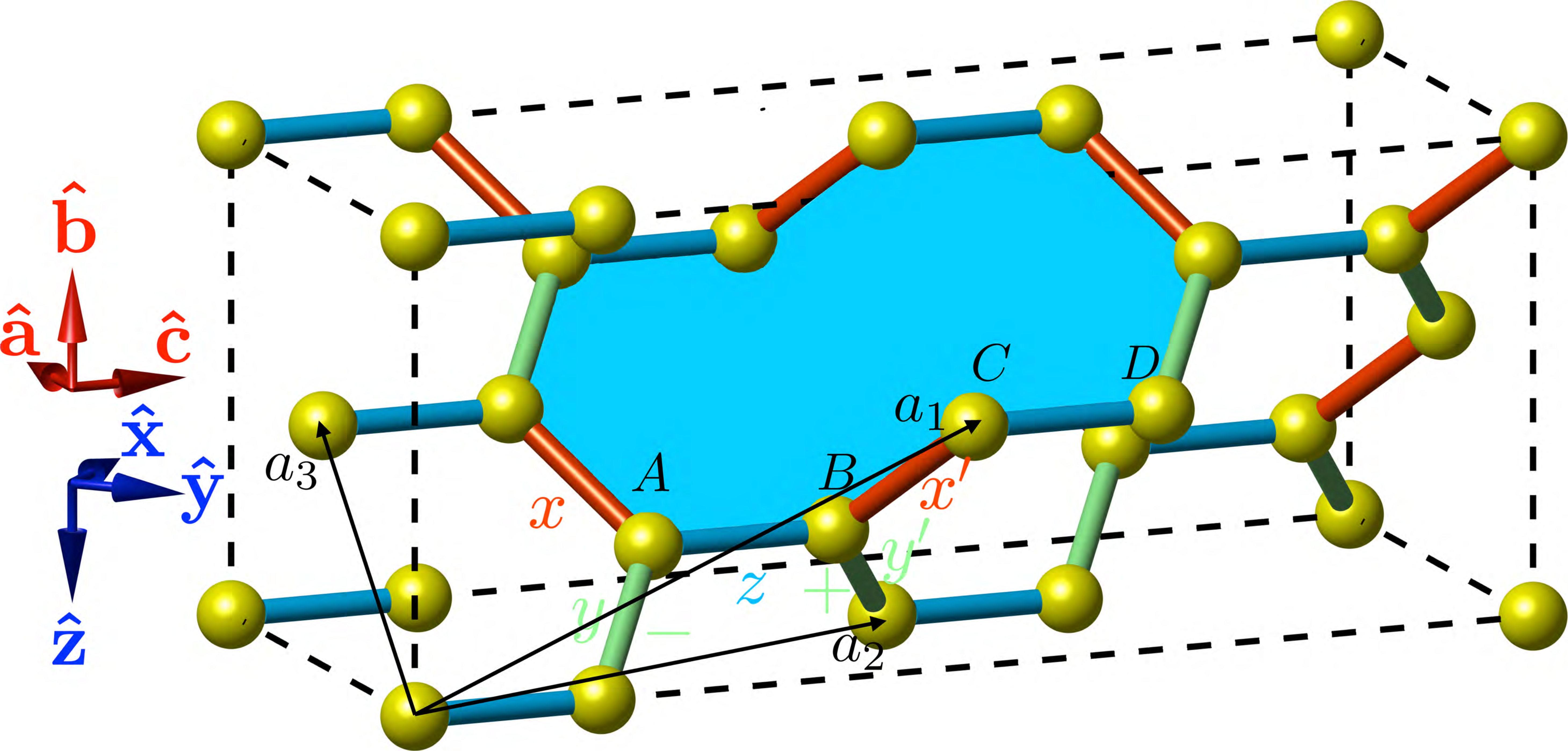}
\caption{The sketch  of the hyperhoneycomb lattice. 
The  conventional orthorhombic unit cell is set by the crystallographic axes $\hat{{\bf a}}, \hat{{\bf b}}$  and $\hat{{\bf c}}$. The three lattice vectors
of the primitive face-centered orthorhombic lattice are given by
$ \mathbf{a}_1 = \left( 0, \sqrt{2}, 3 \right)$, 
$ \mathbf{a}_2 =\left( 1, 0, 3 \right)$, 
$ \mathbf{a}_3 = \left( 1, \sqrt{2}, 0 \right)$ which is written in the crystallographic basis. The four sublattices $A,\, B,\, C$ and $D$ are shown, and we set  $\mathbf{r}_A = \left( 0, 0, 0 \right)$. Different bond types
$x$, $y$, and $z$ are marked by red, green, and blue, respectively.  
The Cartesian axes $\{\hat{{\bf x}}, \hat{{\bf y}}, \hat{{\bf z}}\}$ used to write the the spin Hamiltonian Eq.~(\ref{eq-H}) is related to the crystallographic orthorhombic axes by
$
\hat{{\bf x}}=(\hat{{\bf a}}+\hat{{\bf c}})/\sqrt{2}$,
$\hat{{\bf y}}=(\hat{{\bf c}}-\hat{{\bf a}})/\sqrt{2}$ and $
\hat{{\bf z}}=-\hat{{\bf b}}$. The shaded  region  denotes a loop on the hyperhoneycomb lattice containing 10 sites.  }
		\label{fig:HClattice}
\end{figure}

\section{The spin-phonon model} \label{ModelSec}
 In this section, we 	
introduce the spin-phonon coupled Kitaev  model on the hyperhoneycomb lattice and discuss its  phonon  dynamics.
It is described by the following Hamiltonian:
\begin{align}
\mathcal{H}=\mathcal{H}^{ s}+\mathcal{H}^{\mathrm{ph}}+\mathcal{H}^{\text c}. \label{eq:model}
\end{align} 
The {\it first term} in Eq. (\ref{eq:model}) is  the spin  Hamiltonian. The {\it  second term} 	
 is the bare Hamiltonian for the acoustic phonons. The  {\it  third term}  is  the magnetoelastic coupling.

\subsection{ The Kitaev model on the  hyperhoneycomb lattice}\label{Kitaevmodel}
 
\begin{figure*} 
    \centering
       \includegraphics[width =0.95\textwidth]{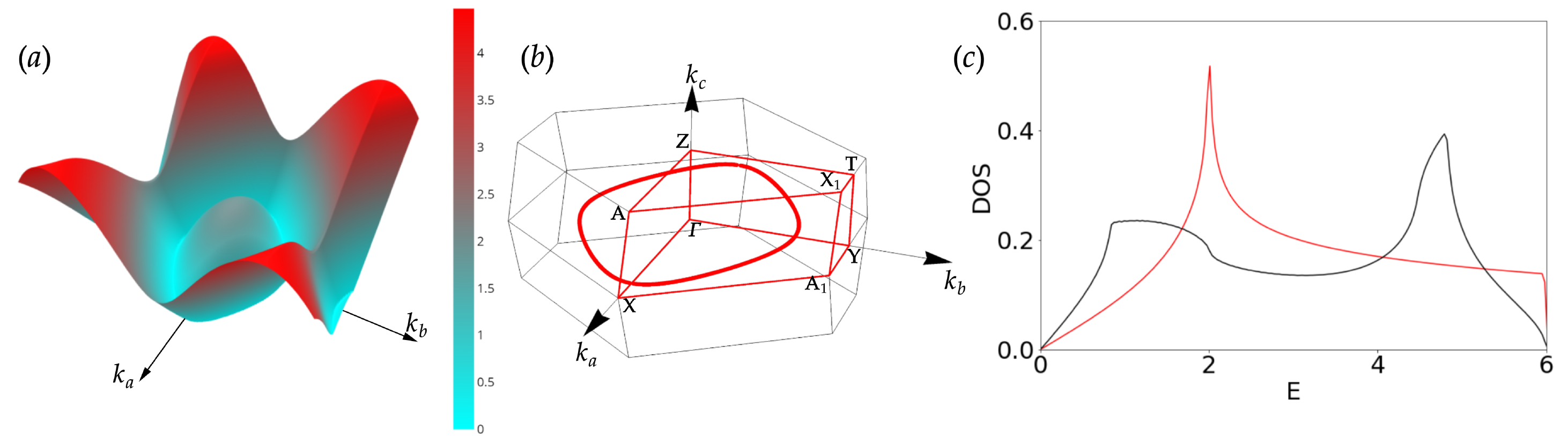}
      \caption{ (a) The dispersion of the lowest branch of the fermionic excitations in the hyperhoneycomb Kitaev model  through the plane of the nodal line ${\bf K}_0=(k_a,k_b,0)$,  whose position in the Brillouin zone  is explicitly shown in panel (b).(c) One-fermion density of states (DOS)  of the isotropic Kitaev models on the  honeycomb (red line) and the hyperhoneycomb (black line) lattices. In each case, the density of states is normalized to unity.}
    \label{fig:spectrum}
\end{figure*}

 We start by  revisiting the  main  features of the Kitaev QSL realized on the hyperhoneycomb 
 lattice previously discussed in Refs.\cite{Hermanns2015,Halasz2017}. 
 The hyperhoneycomb lattice is a face-centered orthorhombic lattice
with four sites per  primitive unit cell. Apart from translational symmetry, the crystal structure is invariant under the $D_{2h}$ point group  symmetry.
 The  conventional orthorhombic unit cell is set by the crystallographic axes $\{\hat{{\bf a}}, \hat{{\bf b}},\hat{{\bf c}}\}$, as shown in \reffg{fig:HClattice}. The Cartesian axes $\{\hat{{\bf x}}, \hat{{\bf y}}, \hat{{\bf z}}\}$ used to write the spin vector field is expressed as
$
\hat{{\bf x}}=(\hat{{\bf a}}+\hat{{\bf c}})/\sqrt{2}$,
$\hat{{\bf y}}=(\hat{{\bf c}}-\hat{{\bf a}})/\sqrt{2}$ and $
\hat{{\bf z}}=-\hat{{\bf b}}$. Different bond types
$x$, $y$, and $z$ are marked by red, green, and blue, respectively. 
Note, however, that there are two non-equivalent types of $x$ and $y$ bonds, and  the
hyperhoneycomb structure can be viewed as a stacking of two types of zigzag chains formed by  $x, y$ bonds and $x', y'$ bonds, each pair running along  ${\bf a}\!+\!{\bf b}$ and ${\bf a}\!-\!{\bf b}$ directions, respectively. The two types of chains are interconnected with vertical $z$-bonds.  Thus, in total, there are five types of  nearest neighboring bonds: $x, x', y, y',$ and $z$.

The Kitaev spin model on the  hyperhoneycomb lattice reads 
\begin{equation}
{\mathcal H}^s \!=\! -J\!\left(
\!
\sum_{\lefta \vrr\vrr'\righta \in 
\{x,x'\}}
\!\!\!\!\! 
\sigma_{\mathbf{r}}^x \sigma_{\mathbf{r}'}^x 
+ 
\!\!\!\!\! 
\sum_{
\lefta \vrr\vrr'\righta \in \{y,y'\}} \!\!\!\!\! 
\sigma_{\mathbf{r}}^y
\sigma_{\mathbf{r}'}^y
+ 
\!\!\!\!\! 
\sum_{\lefta \vrr\vrr'\righta \in \{
z\} } 
\!\!\!\!\! 
\sigma_{\mathbf{r}}^z \sigma_{\mathbf{r}'}^z\right),
\label{eq-H}
\end{equation}
where
$\mathbf{r}$ and $\mathbf{r}'$ are sites on
 the three-dimensional  hyperhoneycomb lattice, which we sketch  in Fig.\ref{fig:HClattice} and the summation is done over five types of bonds.
We also assumed the isotropic case with 
$J_{x}\!\!=\!\!J_{y}\!\!=\!\!J_{z}\!\!=\!\!J$.
The symmetry of the  Hamiltonian (\ref{eq-H})  involves a combined lattice and spin transformations \cite{you2012doping}
[for detailed mathematical description, see Ref.~\cite{feng2022phonon, Kexin2021}].
The results of the three $\pi$-rotations around the crystallographic axes ${\bf a}$,  ${\bf b}$, and  ${\bf c}$ are the following.
 Under $C_{2\bf a}$ rotation spins transform as $[\sigma^x,\sigma^y,\sigma^z]\rightarrow[-\sigma^y,-\sigma^x,-\sigma^z]$, 
 under $C_{2\bf b}$ rotation $[\sigma^x,\sigma^y,\sigma^z]\rightarrow[-\sigma^x,-\sigma^y,\sigma^z]$,  and under $C_{2\bf c}$ rotation $[\sigma^x,\sigma^y,\sigma^z]\rightarrow[\sigma^y,\sigma^x,-\sigma^z]$.
 Additionally, $D_{2h}$ group also contains the space inversion $I$  at the middle of $x,x',y,y'$ bonds, which together with spin transformation leads to $[\sigma^x,\sigma^y,\sigma^z]\rightarrow[\sigma^y,\sigma^x,\sigma^z]$. The transformation $C_{2\bf c}$, $C_{2\bf b}$ and $I$ constitute the canonical generators that generate the whole $D_{2h}$ group.

The exact solution of model  (\ref{eq-H}) is based on the macroscopic number of local symmetries in the 
  products of particular components of the spin operators around every plaquette $P$, which on the hyperhoneycomb lattice consists of ten sites (see shaded region in Fig.\ref{fig:HClattice}) and is defined  by the following plaquette operator
$\hat{W}_p=\prod_{\mathbf{r} \in  P} \sigma_\mathbf{r}^{\alpha (\mathbf{r})}$,
where the spin component $\alpha(\mathbf{r})$ is given by the label of the outgoing bond direction.  Since all plaquette operators $\hat{W}_p$ commute with the Hamiltonian, $[\hat{W}_p, \mathcal{H}_{ s}]=0$, and take eigenvalues of $\pm 1$, the Hilbert space  of  the spin Hamiltonian $\mathcal{H}_{ s}$  can be divided into eigenspaces of $\hat{W}_p$.  The ground state of the Kitaev model on the hyperhoneycomb lattice
 is the  zero-flux state with all $\hat{W}_p=1$ \cite{Mandal2009, Hermanns2015}.  This, however, can not be derived exactly from the Lieb's theorem \cite{Lieb1994} but is only based on the numerical calculations \cite{Hermanns2015,Eschmann2020}. Thus,  strictly speaking, the
Kitaev model on hyperhoneycomb lattice is not exactly solvable. 
 Another striking difference between the hyperhoneycomb Kitaev spin liquid 
 and its two-dimensional counterpart regards the effect of the thermal 
fluctuations on the stability of the ground-state zero-flux state.
 While in two-dimensional honeycomb lattice thermal  fluctuations
immediately destroy the zero-flux order of the $Z_2$ gauge field \cite{Nasu2014,Feng2020}, in
three spatial dimensions  there is a finite-temperature transition separating it
from a high-temperature disordered flux state \cite{Nasu2014b,Yoshitake2017,Eschmann2020}.

Using the Kitaev's representation of spins in terms of Majorana fermions \cite{Kitaev2006},
$\sigma_{\mathbf{r}}^{\gamma} = i b_{\mathbf{r}}^{\gamma}
c_{\mathbf{r}}^{\phantom{\gamma}}$ with $\gamma = x,y,z$ \cite{Kitaev2006},
the spin  Hamiltonian Eq.(\ref{eq-H}) can be  rewritten as
\begin{equation}
\mathcal{H}^s= \sum_{\gamma} \sum_{\langle \mathbf{r}, \mathbf{r}'
\rangle_{\gamma}} i J_{\gamma}^{\phantom{\gamma}} \eta_{\mathbf{r},
\mathbf{r}'}^{\gamma} c_{\mathbf{r}}^{\phantom{\gamma}}
c_{\mathbf{r}'}^{\phantom{\gamma}} = \frac{1}{2} \sum_{\mathbf{r},
\mathbf{r}'} \mathcal{H}_{\mathbf{r}, \mathbf{r}'} c_{\mathbf{r}}
c_{\mathbf{r}'}, \label{eq-H-2}
\end{equation}
where $\eta_{\mathbf{r}, \mathbf{r}'}^{\gamma} \equiv i
b_{\mathbf{r}}^{\gamma} b_{\mathbf{r}'}^{\gamma} = \pm 1$, $\mathcal{H}_{\mathbf{r},
\mathbf{r}'} = i J_{\gamma}^{\phantom{\gamma}} \eta_{\mathbf{r},
\mathbf{r}'}^{\gamma}$ if $\mathbf{r}$ and $\mathbf{r}'$ are
neighboring sites connected by a $\gamma$ bond and
$\mathcal{H}_{\mathbf{r}, \mathbf{r}'} = 0$ otherwise.
In the ground-state 
flux sector, we choose the gauge sector with  all $\eta_{\mathbf{r}, \mathbf{r}'}^{\gamma}=1$, which corresponds to all  $\hat{W}_p=1$.
 The quadratic fermionic Hamiltonian in Eq.~(\ref{eq-H-2}) can be
diagonalized via a standard procedure \cite{Kitaev2006}. Since the hyperhoneycomb  lattice  has four sites per unit cell,
the resulting band structure has four fermion bands,  $\xi=1\sim4$ ($\xi = 1, 2$ are the two positive bands). The diagonal form of the
Hamiltonian \cite{Halasz2017}
\begin{equation}\label{Hs-diag}
{\mathcal H}^s= \sum_{\mathbf{k}} \sum_{\xi = 1}^{4}
\varepsilon_{\mathbf{k}, \xi}^{\phantom{\dag}} [\psi_{\mathbf{k},
\xi}^{\dag} \psi_{\mathbf{k}, \xi}^{\phantom{\dag}} - 1/2]
\end{equation}
 is then obtained by the unitary transformation  $\tilde{\mathcal{H}}_{\mathbf{k}}^{\phantom{\dag}} =
\mathcal{W}_{\mathbf{k}}^{\phantom{\dag}} \cdot
\mathcal{E}_{\mathbf{k}}^{\phantom{\dag}} \cdot
\mathcal{W}_{\mathbf{k}}^{\dag}$ of the Hermitian matrix
$\tilde{\mathcal{H}}_{\mathbf{k}}^{\phantom{\dag}}$ with elements
$
(\tilde{\mathcal{H}}_{\mathbf{k}})_{\nu \nu'} = \frac{1} {N}
\sum_{\mathbf{r} \in \nu} \sum_{\mathbf{r}' \in \nu'}
\mathcal{H}_{\mathbf{r}, \mathbf{r}'} \, e^{i \mathbf{k} \cdot
(\mathbf{r}' - \mathbf{r})}$, where $\nu$ and $\nu'$ denote sublattices $a,b,c,d$ shown in Fig. \ref{fig:HClattice}, and
$\varepsilon_{\mathbf{k}, \xi} = (\mathcal{E}_{\mathbf{k}})_{\xi
\xi}$ are the fermionic energies. The fermionic eigenmodes  are given by
\begin{equation}
\psi_{\mathbf{k}, \xi}^{\phantom{\dag}} = \frac{1} {\sqrt{2N}}
\sum_{\nu = 1}^{n} \left( \mathcal{W}_{\mathbf{k}}^{\dag}
\right)_{\xi \nu} \sum_{\mathbf{r} \in \nu}
c_{\mathbf{r}}^{\phantom{\dag}} \, e^{-i \mathbf{k} \cdot
\mathbf{r}}. \label{eq-psi}
\end{equation}
Note that only the fermions $\psi_{\mathbf{k}, \xi}$ with energies
$\varepsilon_{\mathbf{k}, \xi} > 0$ are physical  due to the
particle-hole redundancy
$\tilde{\mathcal{H}}_{-\mathbf{k}}^{\phantom{\dag}} =
-\tilde{\mathcal{H}}_{\mathbf{k}}^{*}$, which implies
$\psi_{-\mathbf{k}, \xi}^{\phantom{\dag}} = \psi_{\mathbf{k},
\xi}^{\dag}$ and  $\varepsilon_{-\mathbf{k}, \xi}^{\phantom{\dag}} = \varepsilon_{\mathbf{k}, \xi}$. Thus, only two branches have positive spectrum. 
 The lowest branch $\varepsilon_{\mathbf{k}, 1}$ [shown in Fig.~\ref{fig:spectrum} (a)]  exhibits the nodal line on  the $(k_a,k_b)$ plane [Fig.~\ref{fig:spectrum} (b)],  which  is  protected  by projective time-reversal  symmetry \cite{Hermanns2015}.  By solving the equation $\varepsilon_{{\mathbf k},1}=0$, we obtained   the functional form  of the nodal line,
 ${\bf K}_0=(k_a,k_b,0)$, with
\begin{align}\label{eq:nodalline}
		k_b
		&= \frac{1}{\sqrt{2}} \arg \left(1 - 2 \cos k_a \pm i \sqrt{1 + 4 \cos k_a - 2 \cos 2 k_a}\right).
	\end{align}
\noindent The  energy dispersion is linear  if expanded around the nodal line, i.e. each point of the nodal  line represents a Dirac cone.  Importantly, the Fermi velocity  varies along the nodal line and  depends on the direction of the deviation from it, i.e. $v_F=v_F({\bf K}_0, \delta{\bf k})$, where $\delta{\bf k}=(\delta k_a, \delta k_b,\delta k_c)$. As we will see later, the spacial dependence of the Fermi velocity of the low-energy
Majorana fermions will lead to the qualitative difference in the temperature dependence of the sound attenuation coefficient  between the hyperhoneycomb model and the honeycomb Kitaev model \cite{Ye2020,Feng2021}.

To further characterized the  spectrum of Majorana fermions, in
Fig.~\ref{fig:spectrum}(c) we plot  the density of states $\text {DOS}(E) =\sum_{\xi=1,2} \int_{BZ} \delta(E-\varepsilon_{{\bf k},\xi}) d^3{\bf k}$ for  the hyperhoneycomb  Kitaev model (shown by the black line)
where the contributions from  both branches of  Majorana fermions are summed up.
The low-energy  DOS is linear in energy,  which follows directly from the linear low-energy dispersion and  the dimension of the Fermi surface \cite{Brent2015}.   For comparison,  in   Fig.~\ref{fig:spectrum}(c) we also plot the DOS for the honeycomb model (shown by the red line). The differences between the DOS for these two lattices can be
understood in terms of the number of fermionic  bands, one for the honeycomb lattice and two for the hyperhoneycomb lattices, and their  nodal structure - two  Dirac points for the honeycomb lattice and the closed line of Dirac points for the hyperhoneycomb lattice. The former leads to the absence of the Van-Hove singularities and overall more flatten DOS   for the hyperhoneycomb lattice. The latter is responsible for a faster
growth of the hyperhoneycomb DOS  at low energies, which is consistent  with higher dimensionality of the nodal line and enlarged number of low-energy states.\\

\subsection{Acoustic phonons on the  hyperhoneycomb lattice}\label{Sec:phonons}	
 Next we find the spectrum of the acoustic phonons on the hyperhoneycomb lattice.
  The   bare Hamiltonian for
the acoustic phonons 
contains the kinetic and elastic energy,
$
 \mathcal{H}^{\mathrm{ph}}=\mathcal{H}_{kinetic}^{\mathrm{ph}}+\mathcal{H}_{elastic}^{\mathrm{ph}}
$,
where
 $\mathcal{H}_{kinetic}^{\mathrm{ph}}=\sum_{{\bf q},\mu}\frac{\bf {P}_{-\bf q,\mu}\cdot \bf {P}_{\bf q,\mu}}{2\rho\delta_V}$ with  $\bf {P}_{{\bf q},\mu}$ denoting the momentum  of the phonon with polarization $\mu$,  $\delta_V$ is the area enclosed in one unit cell and $\rho$ is the mass density of the lattice ions. The elastic contribution $\mathcal{H}_{elastic}^{\mathrm{ph}}$ can be expressed in terms of the strain tensor $\epsilon_{ij}=\frac{1}{2}\left( \partial_i u_j+\partial_j u_i\right)$, where ${\bf u}=\{u_a,u_b,u_c\}$ describes the displacement of an atom from its original location.

In order to describe the dynamics of the low-energy acoustic phonons,
 it is convenient to move away from the Hamiltonian formulation and employ instead the long-wavelength effective action $\mathcal{S}$ approach. 
To lowest order, it reads \cite{LandauElasticity}
\begin{align}
\mathcal{S}_{\mathrm{ph}}^{(s)}&=\int\diff^2 x\diff \tau \, [\rho\, (\partial_{\tau} {\bf u})^2+F],\,\, F=\frac{1}{2}\mathcal{C}_{ijlk}\epsilon_{ij}\epsilon_{lk},
\end{align}
where $F$ is the elastic free energy and  $C_{ijlk}$ denote the elements of  the elastic modulus tensor.  
 The number of independent non-zero  $\mathcal{C}_{ijlk}$ is dictated by symmetry. The hyperhoneycomb lattice has  Fddd space group, which is
generated by three glide planes, which are passing through
the bond center of  either of $x,x',y,y'$ bonds and are orthogonal to the  ${\bf a},\,{\bf b},\,{\bf c}$ axes, respectively.
The hyperhoneycomb lattice also has inversion symmetry with respect to the bond center of $x,x',y,y'$ bonds. The inversion thus  can be  generated by the product of glide mirrors, e.g., the inversion on the $x$ bond can be generated by  $d_1^{-1}d_2d_3^{-1}$, where  each $d_i$ glide is accompanied by  a half of  lattice translation along the primitive lattice vector ${\bf a}_i$ \cite{Choi2019}.  Thus  the point group is isomorphic   to  the  $D_{2h}$, for which
  there are  nine independent 
 non-zero components of the elastic modulus tensor 
  $C_{iiii}, C_{ijij},C_{iijj}$, where $i$ and $j$  denote $a,b,c$.
Performing the Fourier transform,
${\bf u} ({\bf r})=\frac{1}{\sqrt{N}}\sum_{\bf q} e^{i{\bf q}\cdot{\bf r}}{\bf u}_{\bf q}$, 
the elastic free energy can  be explicitly written as 
\begin{widetext}
\begin{equation}\label{eq:24}
F=	\frac{1}{2}	\begin{pmatrix} 
		C_{aaaa} q_a^2 + C_{acac} q_c^2 + 
		C_{abab} q_b^2 & 
		q_b q_a \, (C_{aabb} + C_{abab})
		& q_a q_c \, (C_{aacc} + C_{acac}) \\
		q_b q_a (C_{aabb}+C_{abab}) & 
		C_{abab} q_a^2 + C_{bcbc} q_c^2 + 
		C_{bbbb} q_b^2 & q_b q_c \, (C_{bbcc} + C_{bcbc}) \\
		q_a q_c \, (C_{aacc} + C_{acac})&
		q_b q_c \, (C_{bbcc}+ C_{bcbc}) &
		C_{acac} q_a^2 + C_{cccc} q_c^2 + 
		C_{bcbc} q_b^2
	\end{pmatrix},
\end{equation}
 \end{widetext}
 where   $q_a = q \sin \theta_\vq \cos \phi_\vq$, $q_b = q \sin \theta_\vq	\sin \phi_\vq$ and
	$q_c = q \cos \theta_\vq$ are the components of the acoustic vector ${\bf q}$ in the orthorhombic reference frame. 
 By  diagonalizing the matrix (\ref{eq:24}), 
 we  compute  eigenmodes,  one  longitudinal and two transverse acoustic modes,  and the corresponding eigenenergies:  the longitudinal and transverse acoustic phonons are then  given by
  \begin{equation}\label{transfo-phonon}
	\begin{aligned}
		\left(\begin{array}{c}
			u_{{\bf q},a}\\
			u_{{\bf q},b}\\
			u_{{\bf q},c}
		\end{array}\right) = 
	\begin{pmatrix}
		R_{11} & R_{12} & R_{13} \\
		R_{21} & R_{22} & R_{23}  \\
		R_{31} & R_{32} & R_{33} 
	\end{pmatrix}
\left(\begin{array}{c}
	{\tilde u}_{{\bf q}}^{\parallel}\\
	{\tilde u}_{{\bf q}}^{\perp_1}\\
	{\tilde u}_{{\bf q}}^{\perp_2}
\end{array}\right),
	\end{aligned}
\end{equation}
where   $\hat{R}\equiv \hat{R} (\theta_\vq, \phi_\vq)$ is the transformation matrix,   ${\tilde u}_{{\bf q}}^{\parallel}$,
$	{\tilde u}_{{\bf q}}^{\perp_1}$ and 
	${\tilde u}_{{\bf q}}^{\perp_2} $   are the longitudinal and transverse acoustic eigenmodes, respectively. 
 The  energies of  the longitudinal and transverse acoustic phonons are
 \begin{align}\label{appeq:PhSpectrum}
 \Omega^{\nu}_{\bf q}=v_s^{\nu}(\theta_\vq,\phi_\vq)q,
\end{align}
where the sound velocities $v_s^{\nu}$ for the longitudinal acoustic mode, $v_s^{\parallel}(\theta_\vq, \phi_\vq)$, and two transverse modes, 
$v_s^{\perp_1} (\theta_\vq, \phi_\vq)$ and $v_s^{\perp_2} (\theta_\vq, \phi_\vq)$ are anisotropic in space.

\begin{figure*} 
\includegraphics[width=0.8\textwidth]{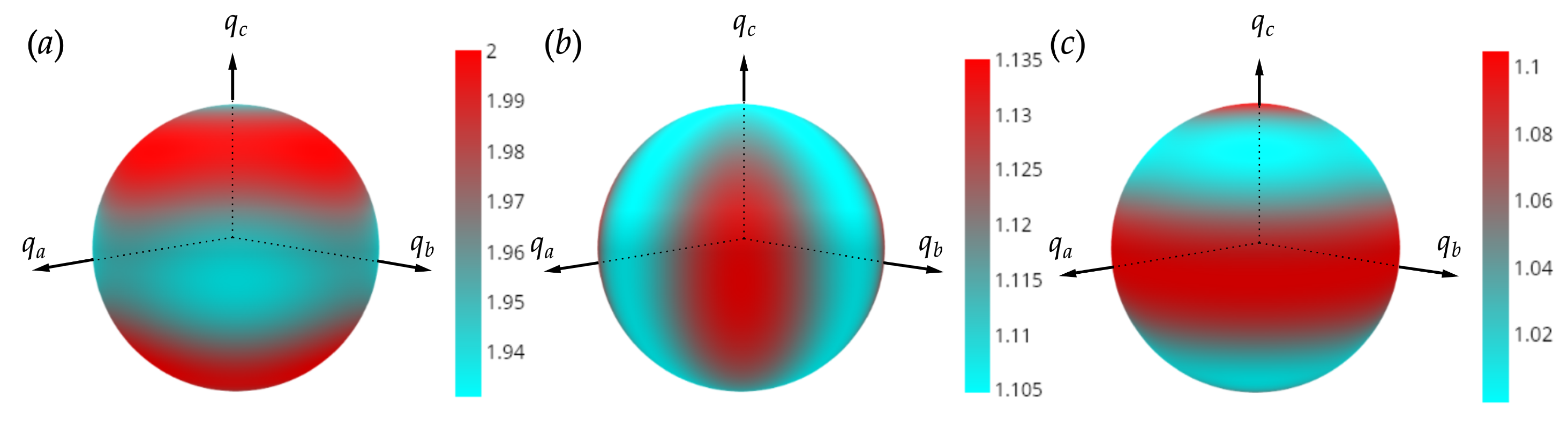}
\caption{ Angular dependence of the sound velocities  [in the unit of  $10^4$ m/s] for (a) longitudinal mode ($v_s^\parallel(\theta, \phi)$) and (b) in-plane transverse mode ($v_s^{\perp_1}(\theta, \phi)$) and (c) out-of-plane transverse mode ($v_s^{\perp_2}(\theta, \phi)$) computed, for the elastic modulus coefficients close to those computed for $\beta$-Li$_2$IrO$_3$ \cite{Tsirlin}.}
\label{fig:velocities}
\end{figure*}

In Fig.~\ref{fig:velocities}, we plot the angular dependence of these velocities computed for the elastic modulus tensor coefficients close to those   computed for  $\beta$-Li$_2$IrO$_3$:
  we set 
 $C_{iiii}=2800$ kbar, $i=a,b,c$, $C_{aacc}=C_{bbcc}=1300$ kbar, $C_{aabb}=C_{abab}=C_{acac}=C_{bcbc}=900$ kbar \cite{Tsirlin}.  We see that the angular dependence of the sound velocities is not that strong.  In the plot, the maximum sound velocity is estimated to be  $2\times 10^4$m/s, which is in the middle  of the sound velocities reported for different directions in $\alpha$-RuCl$_3$  \cite{Lebert2022}.  
For the elastic modulus tensor given above, and restricting phonon modes to $ab$, $bc$ and $ac$ crystallographic planes, 
we numerically
checked that the first column of the rotation matrix  $\hat{R}$ corresponding to the longitudinal mode indeed  gives  the vector parallel to $\vq$, 
i.e.~$\left(R_{11}, R_{21}, R_{31}\right)^\T =  \left(\sin\theta_\vq\cos\phi_\vq, \sin\theta_\vq\sin\phi_\vq, \cos\theta_\vq\right)^\T$,
while the second and the third columns are perpendicular to $\vq$ (the second column with label $\perp_1$ corresponds to the in-plane transverse mode, and the third column with label $\perp_2$  corresponds to the out-of-plane transverse mode).


Knowing the acoustic phonon dispersion relations (\ref{appeq:PhSpectrum}), we can now determine the free phonon propagator in terms of lattice displacement field $\tilde{u}_{{\bf q}}^{\nu}$ as 
\begin{align}
D_{{\bf q}}^{(0)\,\nu\nu'}(t)=-i\langle {\mathcal T} \tilde{u}_{-{\bf q}}^{\nu}(t)\tilde{u}_{{\bf q}}^{\nu'}(0)\rangle^{(0)},
\end{align}
where ${\mathcal T}$ is time ordering operator, the superscript $(0)$ denotes the bare propagator,   $\nu=\parallel,\perp_1, \perp_2$ labels the polarization, and
$ \tilde{u}_{{\bf q}}^{\nu}$ are  phonon eigenmodes in the corresponding polarization, which in the 
second quantized form can be written  as
\begin{align}
 \tilde{u}_{{\bf q}}^{\nu}(t)=i\left(\frac{\hbar}{2\rho\,\delta_V \Omega^\nu_{\bf q}}\right)^{1/2} ({\tilde a}_{\bf q} e^{-i\Omega^\nu_{\bf q} t}+{\tilde a}^{\dagger}_{-{\bf q}}e^{i\Omega^\nu_{\bf q} t}),
\end{align}
where $\delta_V$ is the area enclosed in one unit cell and $\rho$ is the mass density of the lattice ions. 
In the momentum and frequency space, the bare phonon propagator is then given by
\begin{align}
D^{(0)\nu\nu }({\bf q},\Omega)=
-\frac{\hbar}{\rho\,\delta_V}\frac{1}{\Omega^2-
(\Omega^\nu_{\bf q})^2+i0^{+}}.
\end{align}
The dynamics of phonons will be thus described by the decay and scattering of these eigenmodes on low-energy  fractionalized excitations of the Kitaev model, which can be accounted for by the phonon  self-energy $\Pi_{\mathrm{ph}}({\bf q},\Omega)$ \cite{Ye2020}, which for this case we will discuss later in Sec.~\ref{sec:polarizationbubble}.
The renormalized phonon propagator is then given by the Dyson equation
$D({\bf q},\Omega)=\left[
\left(D^{(0)}({\bf q},\Omega)\right)^{-1}-\Pi_{\mathrm{ph}}({\bf q},\Omega)
\right]^{-1}$.


\subsection{The Majorana fermion-phonon coupling vertices}\label{Sec:MFPhcoupling}
\begin{figure}
		\includegraphics[width=.99\linewidth]{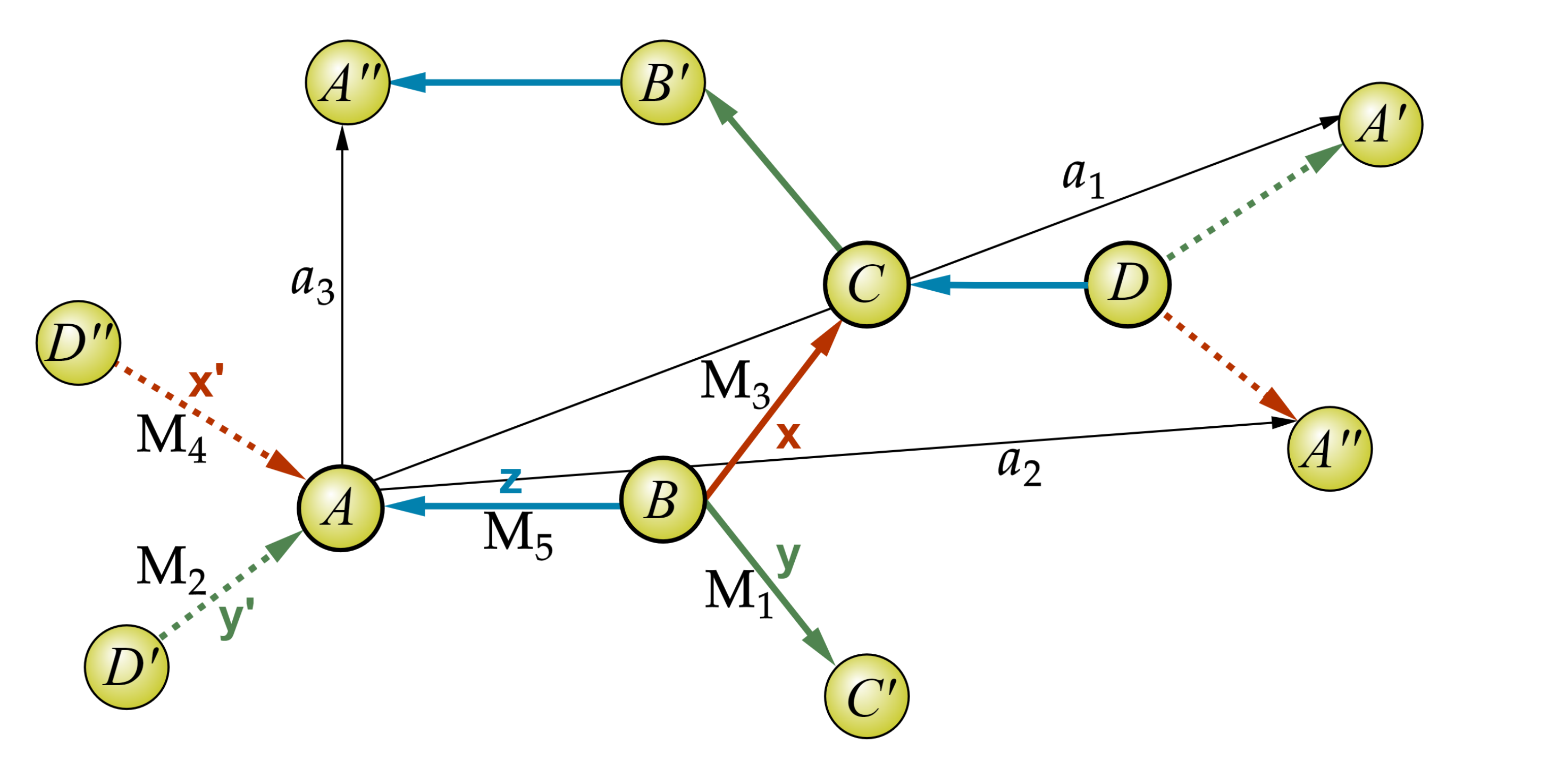}
		\caption{  $A,B,C,D$ denote four sublattices of the hyperhoneycomb lattice. ${\bf M}_1= \frac{1}{2} (1, \sqrt{2}, -1)$, ${\bf M}_2= \frac{1}{2} (1, -\sqrt{2}, -1)$, $	{\bf M}_3= \frac{1}{2} (-1, -\sqrt{2}, -1)$,
		$ {\bf M}_4= \frac{1}{2} (-1, \sqrt{2}, -1)$  and  ${\bf M}_5= (0,0,1)$ are five nearest neighboring  vectors, corresponding to $y,y',x,x',z$ bonds, respectively (all the vectors are given in the crystallographic axes $\hat{{\bf a}}, \hat{{\bf b}}$  and $\hat{{\bf c}}$).  We use the following convention: an arrow pointing from site $\mathbf{r}$ to $\mathbf{r}'$
means $ u_{\mathbf{r}, \mathbf{r}'}$ on the corresponding bond is
positive.  }
		\label{fig:latticebonds}
\end{figure}

In order to study the phonon dynamics in the Kitaev spin liquid, it remains  to compute the Majorana fermion-phonon (MFPh) coupling vertices, which we will do in this section. We recall that the magneto-elastic coupling $\mathcal{H}^{\text c}$ arises from the change in the Kitaev coupling  due to the lattice vibrations.
In the long wavelength limit for acoustic phonons, the coupling Hamiltonian on the bond can be written in a  differential form as
\begin{align}
\label{Cmodel1}
\mathcal{H}^{\text c}_{{\bf r}, {\bf r}+{\bf M}_\alpha}  
= \lambda  {\bf M}_\alpha\cdot \left[ \left(
{\bf M}_\alpha \cdot{\bf \nabla} \right)
{\bf u}({\bf r})\right] \sigma_{\bf r}^{\alpha}  \sigma_{ {\bf r}+{\bf M}_\alpha }^{\alpha},
\end{align}
where $\lambda \sim\left(\frac{\mathrm{d} J}{\mathrm{d} {r}}\right)_\tx{e q} \ell_{a}$  is the strength of the spin-phonon interaction and   $\ell_{a}$ is the lattice constant, and  ${\bf M}_\alpha={\bf M}_1,...
{\bf M}_5$ are five  nearest neighboring  vectors corresponding, respectively, to $y,y',x,x',z$  bonds
shown in Fig.\ref{fig:latticebonds}.
Using these vectors, we can write the spin-phonon coupling Hamiltonian explicitly:
	\begin{widetext}
	\begin{align} \label{eq:3dm}
	\mathcal{H}^{\text c} =&\frac{1}{4}\lambda \sum_{{\bf r}_A}  \bigg( 
	4\sigma_{{\bf r}_A}^z \sigma_{{\bf r}_A+  {\bf M}_5}^z \epsilon_{cc} +
	\sigma_{{\bf r}_A}^y \sigma_{{\bf r}_A +  {\bf M}_2}^y
	 (\epsilon_{aa}+2\epsilon_{bb}+ \epsilon_{cc} - 2\sqrt{2}(\epsilon_{ab}-\epsilon_{bc})-2\epsilon_{ac}) \nonumber\\&+
	  \sigma_{{\bf r}_A}^x\sigma_{{\bf r}_a +  {\bf M}_4}^x
	 (\epsilon_{aa}+2\epsilon_{bb}+ \epsilon_{cc} - 2\sqrt{2}(\epsilon_{ab}+\epsilon_{bc})+2\epsilon_{ac})\bigg) \\& +\sum_{{\bf r}_B}  \bigg( 	
	 \sigma_{{\bf r}_B}^y \sigma_{{\bf r}_B+  {\bf M}_1}^y	 (\epsilon_{aa}+2\epsilon_{bb}+ \epsilon_{cc} +2 \sqrt{2} (\epsilon_{ab}-\epsilon_{bc})-2\epsilon_{ac})		 + \sigma_{{\bf r}_B}^x \sigma_{{\bf r}_B +  {\bf M}_3}^x
	 (\epsilon_{aa}+2\epsilon_{bb}+ \epsilon_{cc} + 2\sqrt{2}( \epsilon_{ab}+\epsilon_{bc})+2\epsilon_{ac})\bigg),\nonumber
\end{align}
\end{widetext}
where we use a short notation $\epsilon_{ij}\equiv \epsilon_{ij} ({\bf r})$ with 
 ${\bf r}={\bf r}_A$ or ${\bf r}_B$ depending on the bond and $i,j$ one of the orthorhombic directions $a,b,c$.

Under the D$_{2h}$ point group symmetry, the spin-phonon Hamiltonian has four independent symmetry channels,
 $A_{g}$, $B_{1g}$, $B_{2g}$,  and $B_{3g}$, which are  inversion-symmetric irreducible representations (IRRs) of this group.
The linear combinations of the strain tensors  that transform as the D$_{2h}$  are  $\epsilon_{aa}$, $\epsilon_{bb}$, and $\epsilon_{cc}$, in the
$A_{g}$ channel, and  $\epsilon_{ab}$, $\epsilon_{ac}$ and $\epsilon_{bc}$ in
 $B_{1g}$, $B_{2g}$,  and $B_{3g}$, respectively. 
 By writing the linear combinations of the Kitaev interactions that transform according  to these IRRs,
 we express
the spin-phonon
coupling Hamiltonian \refeq{eq:3dm} 
as a sum of four independent contributions,
${\mathcal H}^c =  \mathcal{H}^c_{A_{g}} +  \mathcal{H}^c_{B_{1g}} +  \mathcal{H}^c_{B_{2g}} +  \mathcal{H}^c_{B_{3g}}$ with
\begin{widetext}
\begin{align}\label{eq:symm3D}
   \mathcal{H}^c_{A_{g}} &= \lambda_{A_{g}} \sum_{{\bf r}_A,{\bf r}_B}
    \left[ 4\epsilon_{cc} \sigma_{{\bf r}_A}^z \sigma_{{\bf r}_A + {\bf M}_{5}}^z 
    +(\epsilon_{aa} + 2 \epsilon_{bb} + \epsilon_{cc}) 
 	(\sigma_{{\bf r}_B}^y \sigma_{{\bf r}_B + {\bf M}_{1}}^y + 
 	\sigma_{{\bf r}_A}^y \sigma_{{\bf r}_A + {\bf M}_{2}}^y+
 	\sigma_{{\bf r}_B}^x \sigma_{{\bf r}_B +{\bf M}_{3}}^x +
 	\sigma_{{\bf r}_A}^x\sigma_{{\bf r}_A + {\bf M}_{4}}^x )   \right ], \nonumber\\
 	\mathcal{H}^c_{B_{1g}} &= \lambda_{B_{1g}} \sum_{{\bf r}_A,{\bf r}_B}  \epsilon_{ab}   
 	(\sigma_{{\bf r}_B}^y \sigma_{{\bf r}_B + {\bf M}_{1}}^y - 
 	\sigma_{{\bf r}_A}^y \sigma_{{\bf r}_A + {\bf M}_{2}}^y+
 	\sigma_{{\bf r}_B}^x \sigma_{{\bf r}_B + {\bf M}_{3}}^x - 
 	\sigma_{{\bf r}_A}^x\sigma_{{\bf r}_A + {\bf M}_{4}}^x),\nonumber \\
 	\mathcal{H}^c_{B_{2g}} &= \lambda_{B_{2g}} \sum_{{\bf r}_A,{\bf r}_B}
 	\epsilon_{ac}  (-
 		\sigma_{{\bf r}_B}^y \sigma_{{\bf r}_B + {\bf M}_{1}}^y - 
 		\sigma_{{\bf r}_A}^y \sigma_{{\bf r}_A + {\bf M}_{2}}^y +\sigma_{{\bf r}_B}^x \sigma_{{\bf r}_B + {\bf M}_{3}}^x + 
 		\sigma_{{\bf r}_A}^x\sigma_{{\bf r}_A + {\bf M}_{4}}^x), \\\nonumber
 	\mathcal{H}^c_{B_{3g}} &= \lambda_{B_{3g}} \sum_{{\bf r}_A,{\bf r}_B}
 	\epsilon_{bc} (-
 	\sigma_{{\bf r}_B}^y \sigma_{{\bf r}_B + {\bf M}_{1}}^y + 
 	\sigma_{{\bf r}_A}^y \sigma_{{\bf r}_A + {\bf M}_{2}}^y
	 +\sigma_{{\bf r}_B}^x \sigma_{{\bf r}_B + {\bf M}_{3}}^x - 
 	\sigma_{{\bf r}_A}^x\sigma_{{\bf r}_A + {\bf M}_{4}}^x ), 
\end{align}
\end{widetext}
 where we absorbed  numerical prefactors into the definitions of the coupling constants $ \lambda_{A_{g}} , \lambda_{B_{1g}}, \lambda_{B_{2g}}$ and  $\lambda_{B_{3g}}$.

Next,
we express the spin operators  in terms of the Majorana fermions   
and assume the ground state flux sector. Then we perform the  Fourier transformation on both the   strain tensor,
$ \epsilon_{ij}({\bf r})=\frac{1}{\sqrt{N}} \sum_{\bf q}\frac{i}{2}\left(q_{i} u_{{\bf q},j}+q_{j} u_{{\bf q},i}\right) e^{i\vq \cdot \vrr}$,  and the
Majorana fermions,
$c_{{\bf r},\alpha}=\sqrt{\frac{2}{N}}\sum_{\bf k} c_{{\bf k},\alpha}e^{i \bf{k}\cdot \bf{r}_\alpha}$, where $\alpha=A,\,B,\, C,\,D$ is the  sublattice label  [see \reffg{fig:latticebonds}].
Now the products of the spin variables on all non-equivalent bonds can be written as (with the long wavelength approximation $\vq\to 0$ applied):
{\small\begin{equation}\label{matrices}\nonumber
	\begin{aligned}
		&\sigma_{\bf{r}}^y \sigma_{\bf{r + M_{1}}}^y \to {\bf A}^\T_{\bf{-q-k}}S_{\bf{k}}^{\dagger}
		\begin{pmatrix}
			0 & -i e^{i \bf{k \cdot a_3}} & 0 & 0\\
			i e^{-i \bf{k \cdot a_3}} & 0 & 0 & 0\\
			0 & 0 & 0 & 0\\
			0 & 0 & 0 & 0
		\end{pmatrix} S_{\bf{k}} \bf{A_{\bf{k}}} , \\
&	\sigma_{\bf{r}}^y \sigma_{\bf{r + M_{2}}}^y \to {\bf A}^\T_{\bf{-q-k}} S_{\bf{k}}^{\dagger}
	\begin{pmatrix}
		0 & 0 & 0 & 0\\
		0 & 0 & 0 & 0\\
		0 & 0 & 0 & i e^{i \bf{k \cdot a_1}}\\
		0 & 0 & -i e^{-i \bf{k \cdot a_1}} & 0
	\end{pmatrix} S_{\bf{k}}\bf{A_{\bf{k}}}, \\
	&\sigma_{\bf{r}}^x \sigma_{\bf{r + M_{3}}}^x \to {\bf A}^\T_{\bf{-q-k}} S_{\bf{k}}^{\dagger}
	\begin{pmatrix}
		0 & -i & 0 & 0\\
		i & 0 & 0 & 0\\
		0 & 0 & 0 & 0\\
		0 & 0 & 0 & 0
	\end{pmatrix}  S_{\bf{k}}\bf{A_{\bf{k}}},\\
&	\sigma_{\bf{r}}^x \sigma_{\bf{r + M_{4}}}^x \to {\bf A}^\T_{\bf{-q-k}}S_{\bf{k}}^{\dagger}
	 \begin{pmatrix}
		0 & 0 & 0 & 0\\
		0 & 0 & 0 & 0\\
		0 & 0 & 0 & i e^{i \bf{k \cdot a_2}}\\
		0 & 0 & -i e^{-i \bf{k \cdot a_2}} & 0
	\end{pmatrix} S_{\bf{k}} \bf{A_{\bf{k}}}, \\ 
	&\sigma_{\bf{r}}^z \sigma_{\bf{r + M_{5}}}^z \to  {\bf A}^\T_{\bf{-q-k}}S_{\bf{k}}^{\dagger}
	\begin{pmatrix}
		0 & 0 & -i & 0\\
		0 & 0 & 0 & i\\
		i & 0 & 0 & 0\\
		0 & -i & 0 & 0
	\end{pmatrix}S_{\bf{k}}  \bf{A_{\bf{k}}},
	\end{aligned}
\end{equation}}
where  ${\bf a}_i$ are the primitive unit vectors, $S_{\bf{k}} = {\rm diag}\{ e^{i \bf{k} \cdot {\bf r}_\alpha}\}_{\alpha = C,B,D,A}$ is the diagonal matrix in the sublattice basis, and 
the vector ${\bf A}_{\bf{k}} = \left(		c_{{\bf k},C} ,
c_{{\bf k},B} ,
c_{{\bf k},D} ,
c_{{\bf k},A}\right)^\T$. The  Majorana-phonon coupling Hamiltonian in the momentum space is now can be written as 
 $
{ \mathcal H}^c = \sqrt{\frac{2}{N}}
\sum_{{\bf q}, {\bf k}} { \mathcal H}_{{\bf q}, {\bf k}}$, where each  contribution ${ \mathcal H}_{{\bf q}, {\bf k}}$ can be decomposed into the irreducible representations $ A_{g} ,B_{1g}, B_{2g}$ and  $B_{3g}$
[see \refapp{app: coupling-details} for explicit expressions]. Note  also that  ${\bf A}_{\bf{k}}$
is written in this particular permuted basis of  the Majorana fermions in order 
to use  the  convenience of the  auxiliary Pauli matrices in the representation of the  coupling Hamiltonians  as shown in \refeq{Hqk}.

 Next we express the phonon modes 
in terms of the transverse and longitudinal eigenmodes  defined  in \refeq{transfo-phonon}.
  Then ${ \mathcal H}_{{\bf q}, {\bf k}}$ terms
in the corresponding polarizations are given by
\begin{align}\label{Hcpolarizations}
	{ \mathcal H}^{\parallel}_{{\bf q},\bf{k}}& = \tilde{u}_{{\bf q}}^{ \parallel} \, {\bf A}_{\bf{-q-k}}^\T \, S_{\bf{k}}^{\dagger} \, \hat{\lambda}_{{\bf q},{\bf k}}^{\parallel} \, S_{\bf{k}} \, {\bf A}_{\bf{k}} \nonumber\\
	{ \mathcal H}^{\perp_1}_{{\bf q},\bf{k}} &= \tilde{u}_{{\bf q}}^{\perp_1} \, {\bf A}_{\bf{-q-k}}^\T \, S_{\bf{k}}^{\dagger} \,  \hat{\lambda}_{{\bf q},{\bf k}}^{\perp_1} \, S_{\bf{k}} \, {\bf A}_{\bf{k}} \\\nonumber
	{\mathcal H}^{\perp_2}_{{\bf q},\bf{k}} &=\tilde{u}_{{\bf q}}^{\perp_1}  \, {\bf A}_{\bf{-q-k}}^\T \, S_{\bf{k}}^{\dagger} \,  \hat{\lambda}_{{\bf q},{\bf k}}^{\perp_2} \, S_{\bf k} \, {\bf A}_{\bf k}.
\end{align}
The explicit expressions for the  MFPh coupling vertices  for longitudinal and transverse phonon modes
are  given in 
\refapp{app: coupling-details}.
Note also  that since we are using the long wavelength limit  for the phonons, we only
 kept   the  leading in $q$ terms in all $\hat{\lambda}_{{\bf q},{\bf k}}^{\nu}$.

\section{Phonon polarization bubble}\label{sec:polarizationbubble}
At the lowest order, the phonon self-energy 
is given by the polarization bubble  \cite{Ye2020}
\begin{align}
\Pi_{\mathrm{ph}}^{\mu\nu}({\bf q},\Omega)
=i\tr{[\hat\lambda^{\mu}_{{\bf q}, {\bf k}}\mathcal{G}({\bf k},\omega)\hat\lambda^{\nu}_{ {\bf q}, {\bf k}}\mathcal{G}({\bf k}-{\bf q},\omega+\Omega)]},
\label{phononbubble}
\end{align}
where $\hat\lambda^{{\mu}(\nu)}_{{\bf q}, {\bf k}}$ are the MFPh coupling vertices for $\mu (\nu)=\parallel,\perp_1,\perp_2$,    and $\mathcal{G}({\bf k},\omega)$ denotes the Majorana fermions Green's function for the lowest fermionic  branch given by 
$
\hat{\mathcal{G}}({\bf k},\omega)=-i\int_{-\infty}^{+\infty}\diff t\langle {\mathcal T} \psi_{{\bf k},1}(t) \psi_{-{\bf k},1}^\T(0)\rangle e^{i\omega t}$ (in the following, we omit the branch index and simply write  $\psi_{{\bf k}}\equiv\psi_{{\bf k},1}$).

Since we are interested in the phonon decay and scattering at finite temperature, 
it is convenient to use the Matsubara representation for the Majorana Green's functions:
\begin{widetext}
\begin{align}
    &\Pi_{\mathrm{ph}}^{\mu \nu}\left(\vq, i \Omega_{n}\right)=
    \int_{\mathrm{BZ}} \mathrm{d} \vk  \operatorname{Tr}\left[
        \hat{\lambda}_{\vq, \vk}^{\mu} \hat{\mathcal{G}}\left(\vk, i \omega_{m}+i\Omega_{n}\right) \hat{\lambda}_{\vq, \vk}^{\nu} \hat{\mathcal{G}}\left(\vk+\vq,  i\omega_{m}\right)
    \right]
   \nonumber \\
    =
    &
    \int_{\mathrm{BZ}} \mathrm{d} \vk  \operatorname{Tr}\left[
        \hat{\lambda}_{\vq, \vk}^{\mu} \hat{\mathcal{W}}_{\vk}^{\dagger} \hat{G}\left(\vk, i\omega_m + i\Omega_n\right) \hat{\mathcal{W}}_{\vk} \hat{\lambda}_{\vq, \vk}^{\nu} \hat{\mathcal{W}}_{\vk+\vq}^{\dagger} \hat{G}\left(\vk+\vq, i\omega_m\right) \hat{\mathcal{W}}_{\vk+\vq}
    \right]
    \\\nonumber
    =&
    \int_{\mathrm{BZ}} \mathrm{d} \vk \sum_{ij} \sum_{l}\left[
        \left(\hat{\mathcal{W}}_{\vk+\vq}\hat{\lambda}_{\vq, \vk}^{\mu} \hat{\mathcal{W}}_{\vk}^{\dagger}\right)
        \hat{E}_{i}
        \left(\hat{\mathcal{W}}_{\vk} \hat{\lambda}_{\vq, \vk}^{\nu} \hat{\mathcal{W}}_{\vk+\vq}^{\dagger}\right)
        \hat{E}_{j}
    \right]_{ll} P_{\vk, ij} \label{eq: bubble}
\end{align}
\end{widetext}
where $\Tr[\ldots]$ in the first two lines sums over the Matsubara frequencies $i \omega_{m}$. $\hat{G}(\vk, \omega) = 
\left(
    \begin{array}{cc}
    g(\vk, \omega) & 0  \\ 
    0 & \bar{g}(\vk, \omega)
    \end{array}
\right)$ is the  quasiparticle Green's function, and
\begin{align} 
g(\vk, \omega) = -i\int\ud t \left\langle {\mathcal T} \psi_{\vk}(t) \psi_{\vk}^{\dagger}(0)\right\rangle e^{i\omega t}=\frac{1}{\omega-\varepsilon_{\vk}+i 0^{+}} \nn
\\
\bar{g}(\vk, \omega) = -i\int\ud t \left\langle {\mathcal T} \psi_{\vk}^{\dagger}(t) \psi_{\vk}(0)\right\rangle e^{i\omega t}=\frac{1}{\omega+\varepsilon_{\vk} -i 0^{+}}.
\end{align}
$\mathcal{W}_\vk$ is the unitary matrix that diagonalizes  the   Majorana fermion Hamiltonian \cite{Halasz2017}.
$\hat{E}_{i} = \left(
    \begin{array}{cc}
    \delta_{i1} & 0  \\ 
    0 & \delta_{i2}
    \end{array}
\right) $ 
serves to pick up a specific entry of a matrix. The summation over the Matsubara frequencies gives the  dynamic part of the matrix entries
\begin{align}
    P_{\vk,ij} = T \sum_{i\omega_{m}}\left[\hat{G}\left(\vk, i\omega_m + i\Omega_n\right)_{ii} \hat{G}\left(\vk+\vq, i\omega_m\right)_{jj}
    \right].
\end{align}
Their explicit expressions are given in \refapp{app:dynamicalfactor}.


\section{Angular dependence of the attenuation coefficient}\label{Sec:atten}

In this section, 
we will compute the attenuation coefficient for 
 the lossy acoustic wave function  which  
decays with distance away from the driving source as
\begin{align}
\label{eq:WaveFunc}
\ve u(\ve x,t)=\ve u_0 e^{-\alpha_s (\ve q) x}e^{i(\Omega_\vq t- \ve q\cdot \ve x)}.
\end{align}
where ${\ve u}(\ve x,t)$ is the  lattice displacement vector, $\ve u_0 = {\ve u}(\ve x ={\ve 0},t=0)$, $\Omega_\vq$  is the acoustic wave frequency and ${\bf q}= q (\sin \theta_\vq \cos \phi_\vq, \sin \theta_\vq \sin \phi_\vq,\cos \theta_\vq)$ is the propagation vector. The attenuation coefficient $\alpha_s^\mu (\ve q)$ for a given phonon polarization $\mu=\parallel,\perp_1,\perp_2$, defined as the inverse of the phonon mean free path, can be  calculated from the diagonal component of the imaginary part of the  phonon self-energy as  \cite{Ye2020} 
\begin{align}
\alpha_s^\mu ({\bf q}) = -\frac{1}{2 \rho \delta_V \left[v_s^{\mu}(\theta_\vq,\phi_\vq)\right]^2
q} {\rm Im} \left. \Pi^{\mu\mu}_{\mathrm{ph}}({\bf q},\Omega)\right|_{\Omega=v_s^\mu(\theta_\vq, \phi_\vq) q}.
\label{eq:Attenuation}
\end{align}

  \begin{figure*}[!t]
	\centering
	\includegraphics[width=0.3\textwidth]{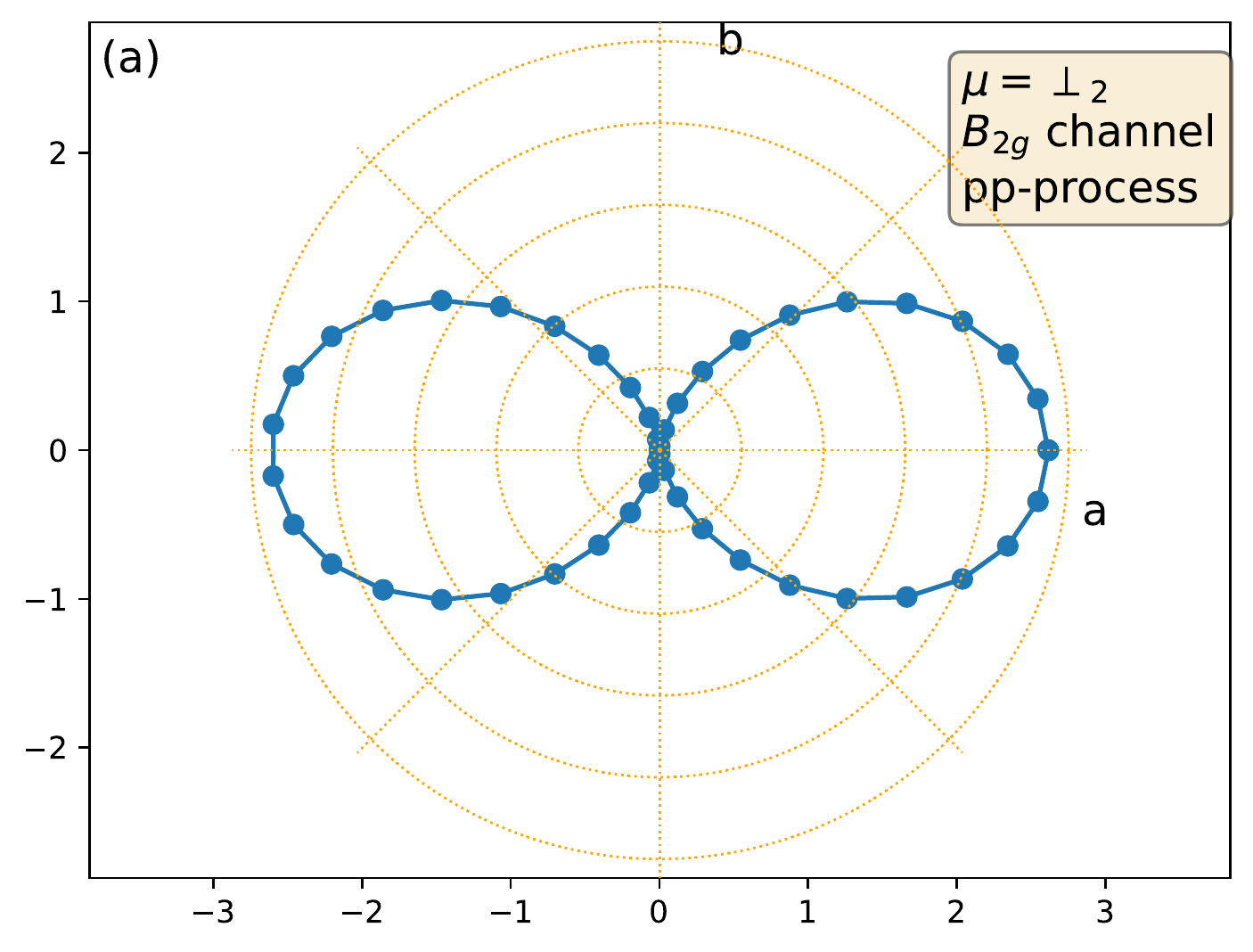}
	\includegraphics[width=0.31\textwidth]{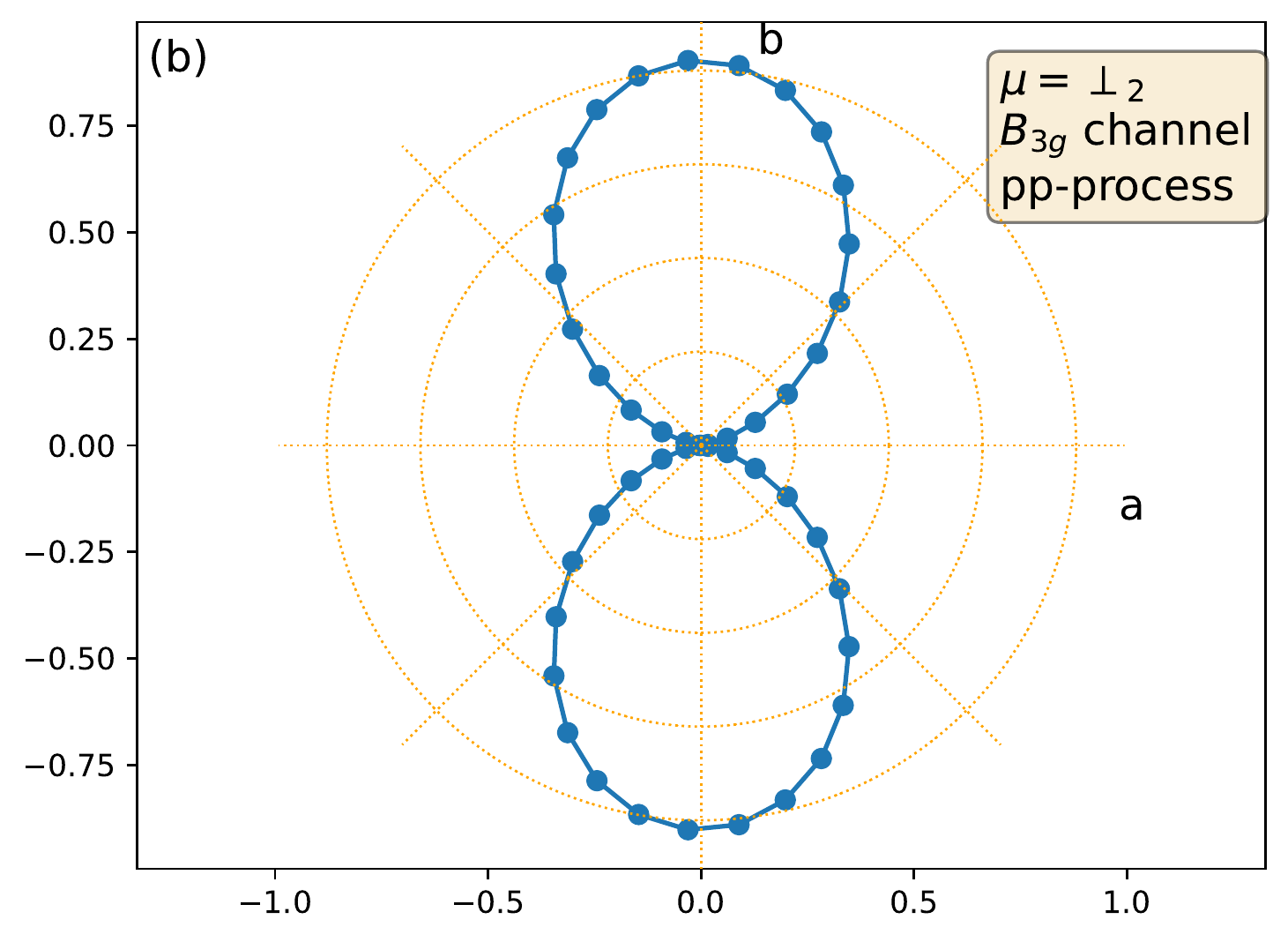}
	\includegraphics[width=0.3\textwidth] {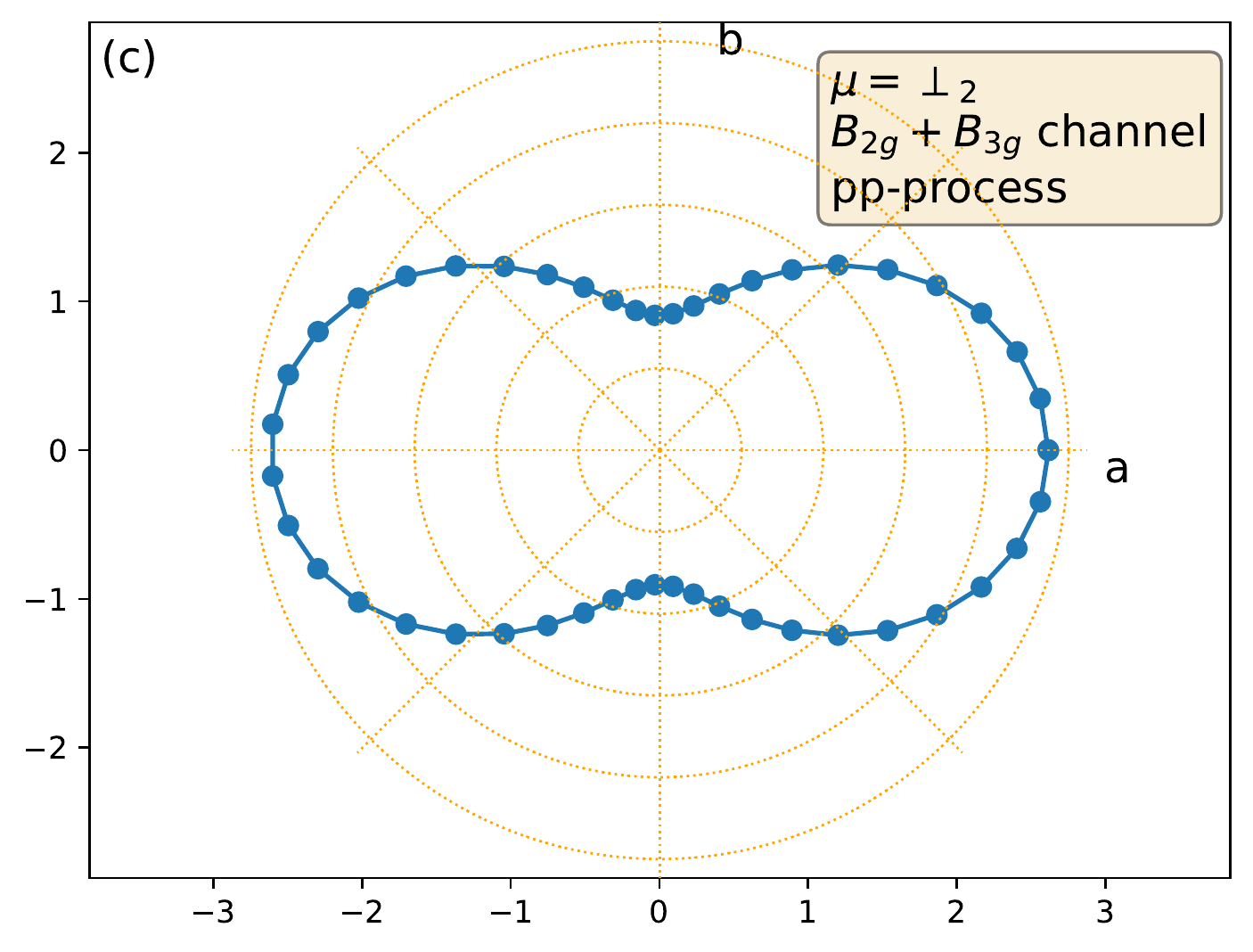}
	\includegraphics[width=0.3\textwidth]{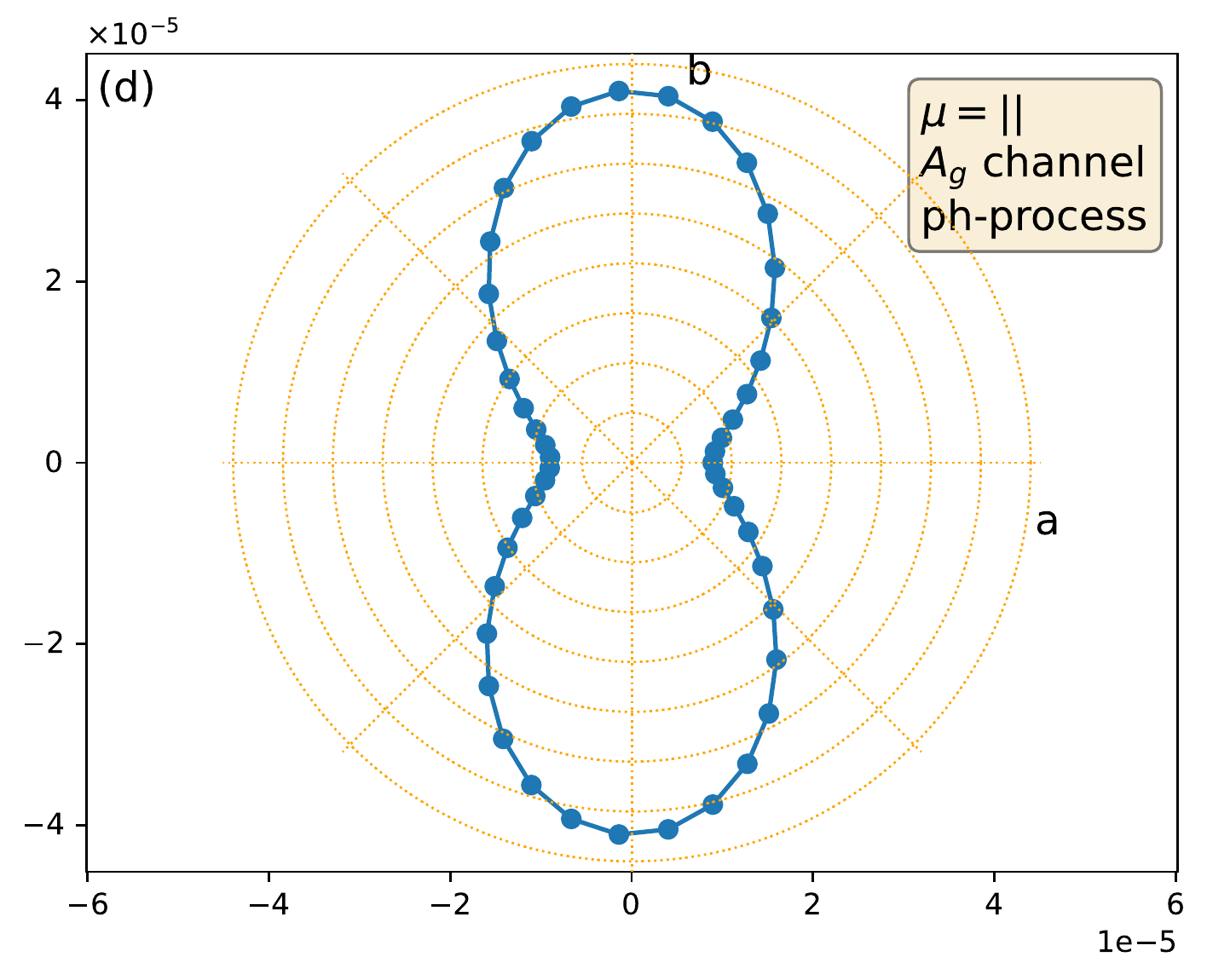}
	\includegraphics[width=0.3\textwidth]{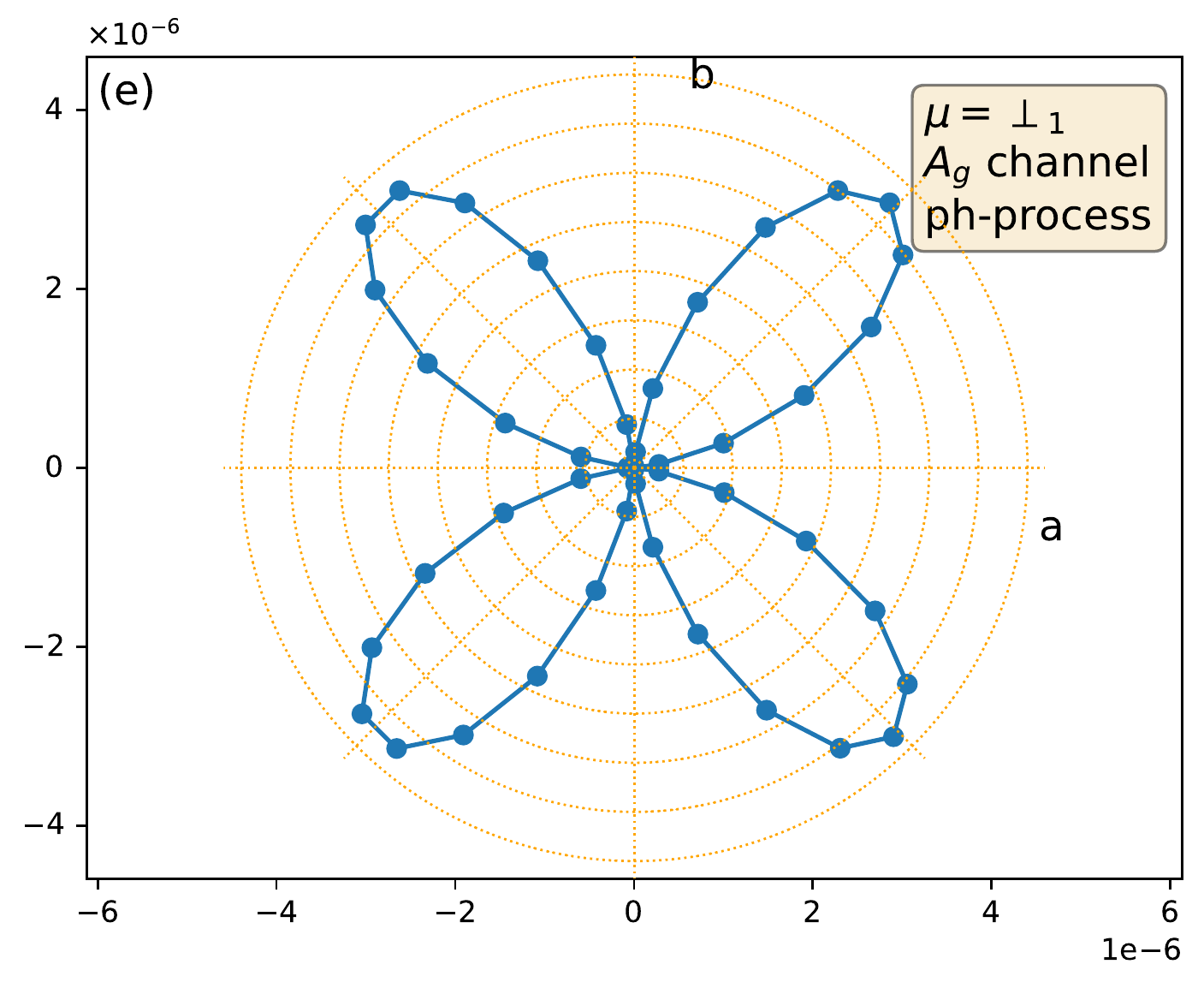}
	\caption{The angular dependence of sound attenuation coefficient $\alpha_s^\mu(\theta_\vq\!=\!90^{\circ}, \phi_\vq)$ in the $ab$-plane. The pp-processes contribute to the attenuation  of  the  out-of-plane transverse phonon mode ($\alpha_s^{\perp_2}$)
	in (a) $B_{2g}$ channel, (b) $B_{3g}$ channel, and (c) combined $B_{2g}$ and $B_{3g}$ channels. The ph-processes contribute to  (d) the attenuation   of the  longitudinal phonon  ($\alpha_s^{\|}$) and  to (e)   the attenuation  of the in-plane transverse phonon mode ($\alpha_s^{\perp_1}$)  in the $A_g$ channel. 
	The radius represents the magnitude of $\alpha_s^\mu$ in the units of $10^{-5} \rho\delta_V$. The calculation is performed  at $T=0.02\, J$.
	}
	\label{fig: ab_pp_ph}
\end{figure*}



\subsection{Kinematic constraints and the estimates for the sound and Fermi velocities in $\beta$-Li$_2$IrO$_3$}\label{Sec:kinematic}

Before analyzing the angular dependence of the sound attenuation coefficient, 
we need first  discuss the  kinematic constraints determining type of the processes involved in sound attenuation. In the  zero-flux low temperature phase, both  momentum and energy are conserved and  kinematic constrains are primarily determined by the relative strength of acoustic phonon velocity $v_s(\theta_\vq, \phi_\vq)$ and Fermi velocity  $v_F({\bf K_0}, \theta_{\delta \vk},\phi_{\delta_\vk})$ (the slope of the Dirac cone at each point of the nodal line), which  in the  most general case are both angular dependent.
These constraints determine whether the decay of the acoustic phonon happens in the particle-
hole (ph-) or in the particle-particle (pp-) channel. Here, by particle and hole, we mean
if the state  of the Majorana fermion  at  $\varepsilon_{\mathbf{k}}$ ($\varepsilon_{\mathbf{k}}=\varepsilon_{\mathbf{k},1}$) is occupied or empty. In other words, the
particle number refers to that of the complex fermion  $\psi_{\mathbf{k}}$ ($\psi_{\mathbf{k}}=\psi_{\mathbf{k},1}$) in Eq.(\ref{Hs-diag}).

 Here we assume that the angular dependence of  $v_s(\theta_\vq, \phi_\vq)$ in  $\beta$-Li$_2$IrO$_3$ is weak [see the  magnitude scale bars in Fig.~\ref{fig:velocities}] and  consider it to be equal to  $v_s$.
However, the Fermi velocity  $v_F({\bf K_0}, \theta_{\delta \vk},\phi_{\delta \vk})$ varies strongly between $v_F=0$ along the nodal line  and  $\max(v_F)$, which can be estimated  from  the  magnitude of the Kitaev coupling, which  in $\beta$-Li$_2$IrO$_3$ is $J\simeq 20\,meV$ \cite{Ruiz2021,Yang2022,Halloran2022}. Taking the lattice constant to be equal to  $\ell = 0.23\, nm$ \cite{Villars2016:sm_isp_sd_1146563}, we estimate $\max(v_F) = 3\, J \ell = 2.1 \times 10^4\, m/s$.
 According to the estimation of the sound velocity in Sec.~\ref{Sec:phonons}, $v_s^{\|} \simeq 2.0 \times10^4 m/s \lessapprox \max(v_F)$, and $v_s^{\perp_{1,2}} \simeq 1.1 \times 10^4 m/s < \max(v_F)$. When  $v_s < \max(v_F)$, the ph-processes are allowed but since they
 require finite occupation number, they scale with $T$ at low temperatures. However, due to the existence of the nodal line along which the Fermi velocity $v_F=0$, the pp-processes are always allowed \cite{Feng2021}. Since they do not require finite occupation number, they are nonzero  even at zero temperature. Therefore, both the pp-processes and ph-processes should be included into  consideration.

\begin{figure*}[!t]
	\centering
	\includegraphics[width=0.3\textwidth]{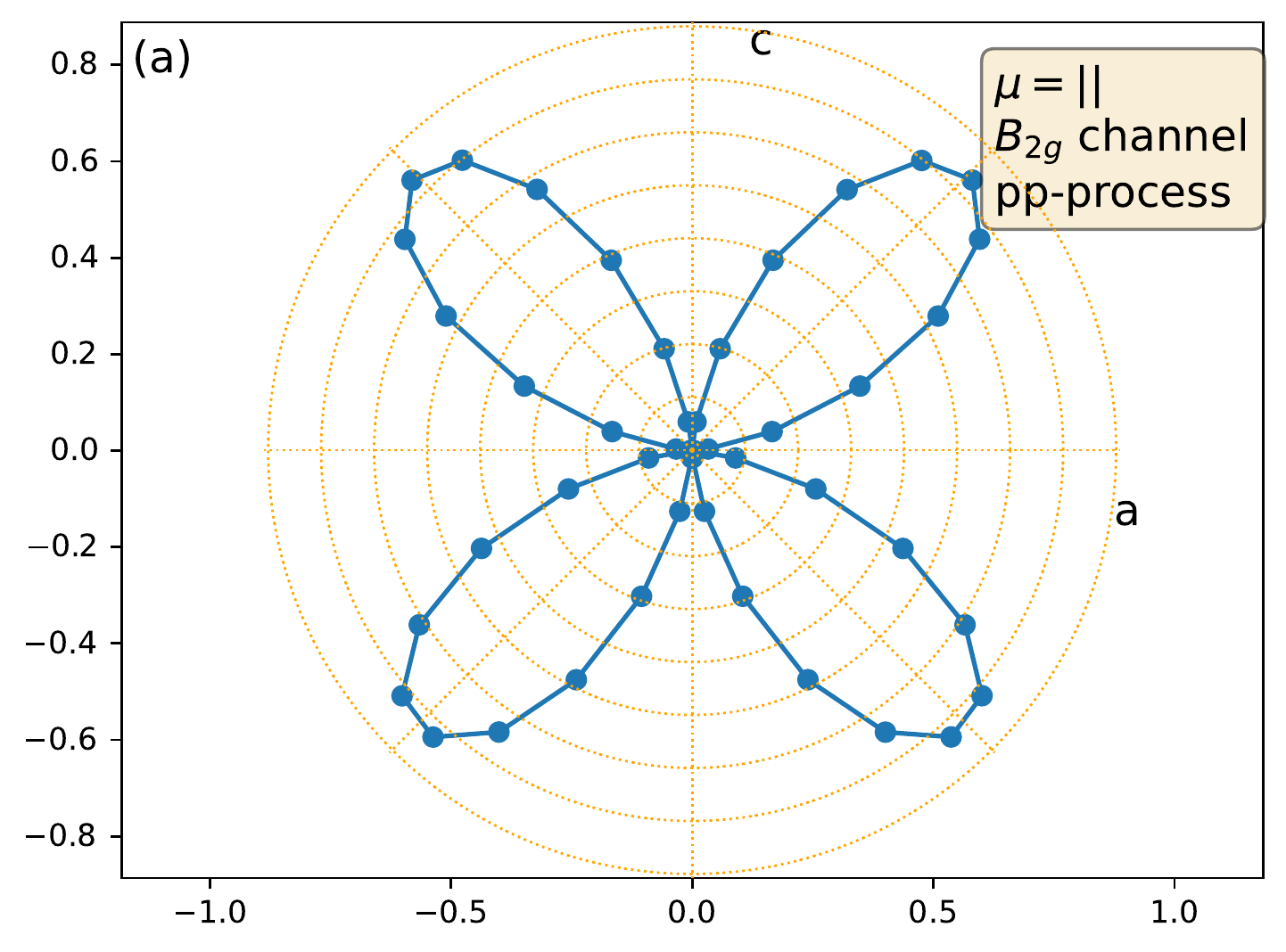}
	\includegraphics[width=0.31\textwidth]{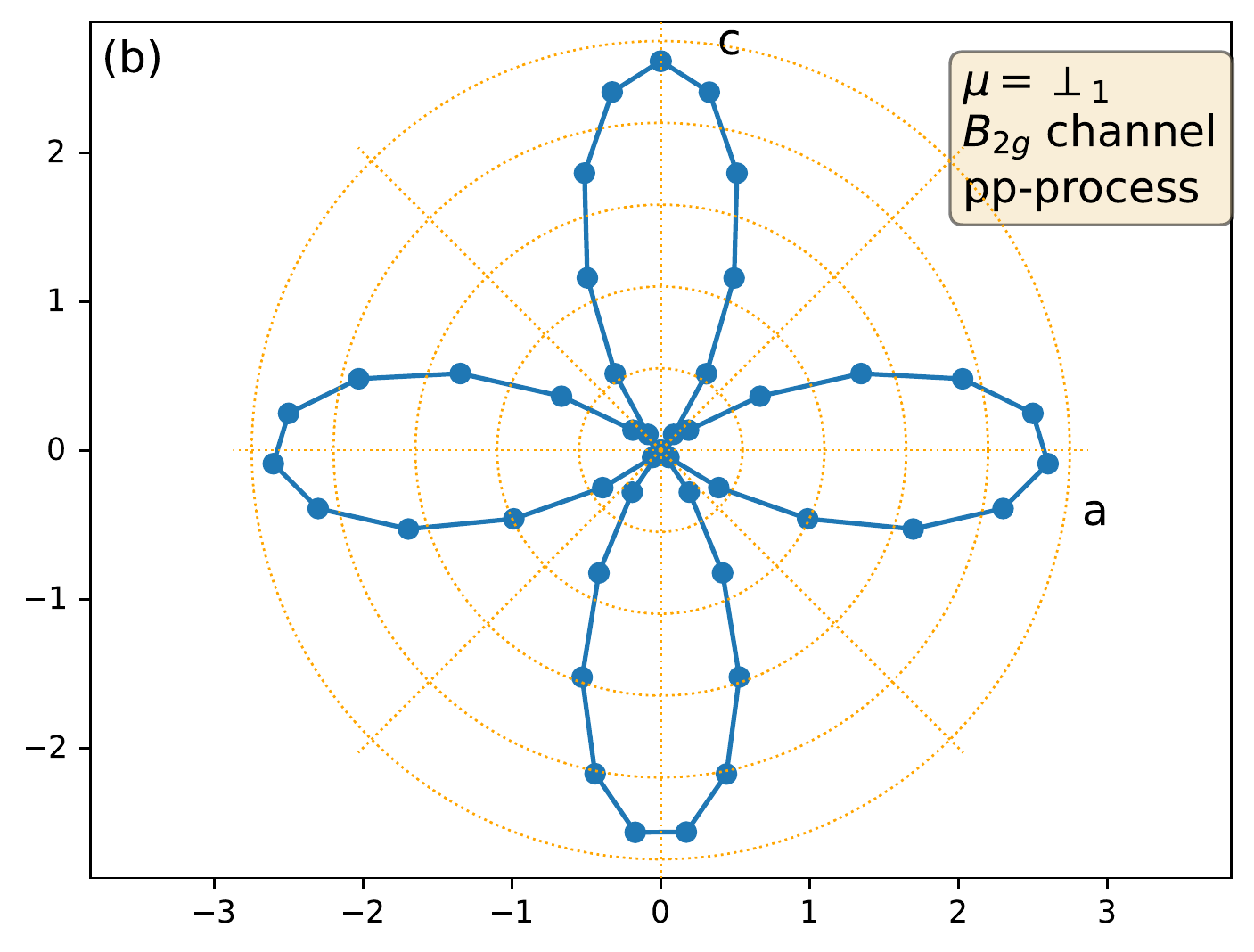}
	\includegraphics[width=0.3\textwidth] {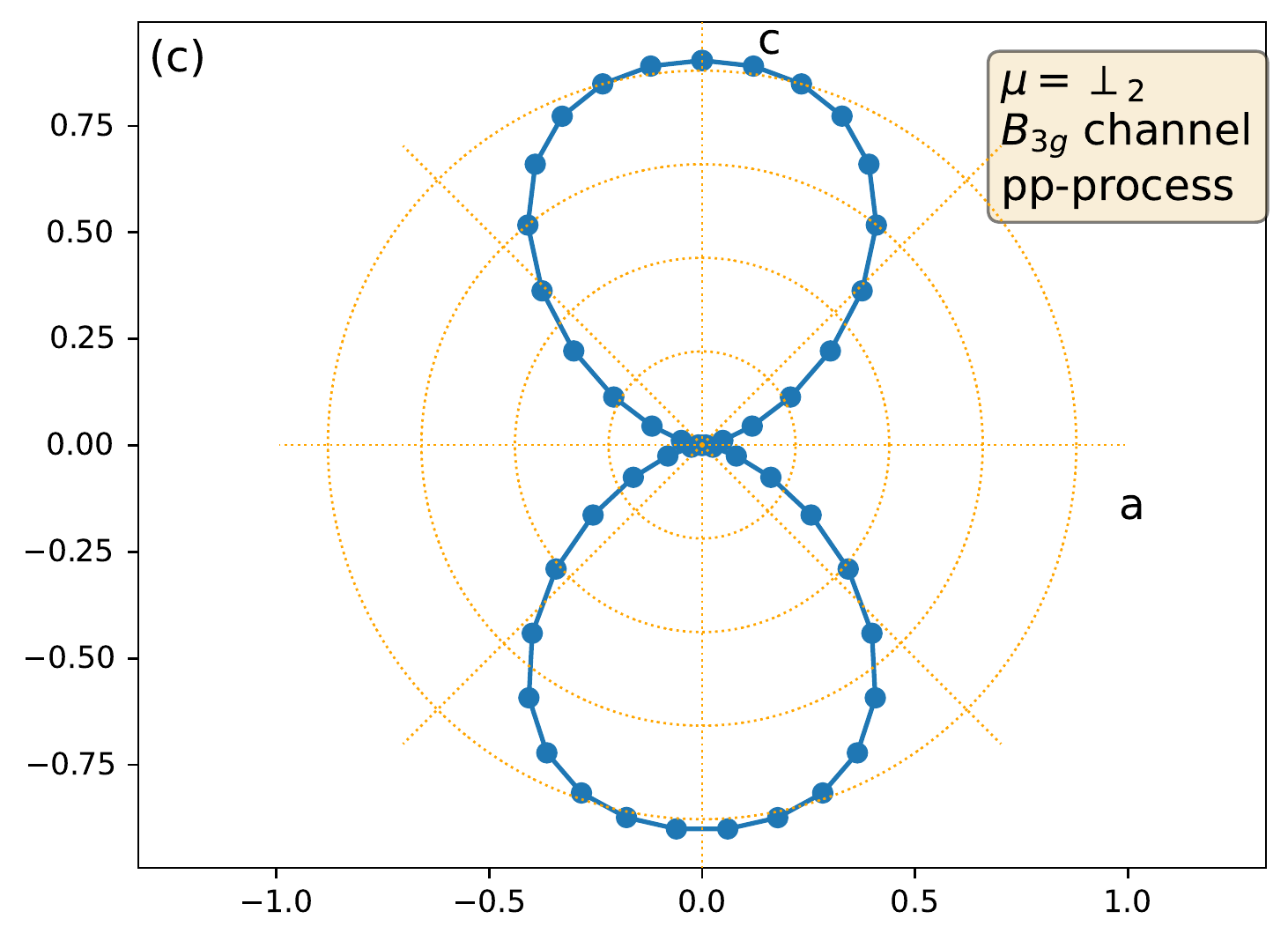}
	\includegraphics[width=0.3\textwidth]{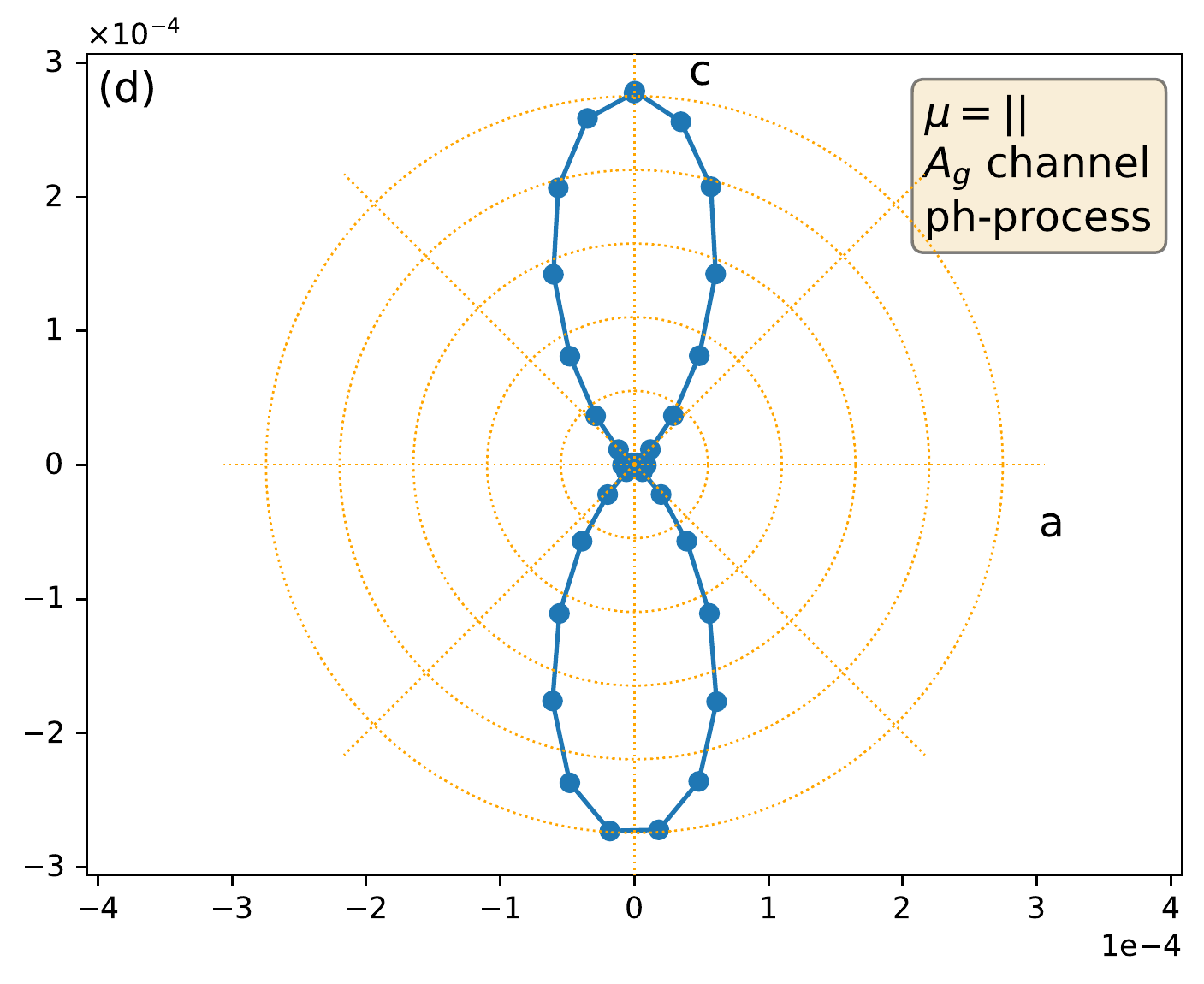}
	\includegraphics[width=0.3\textwidth]{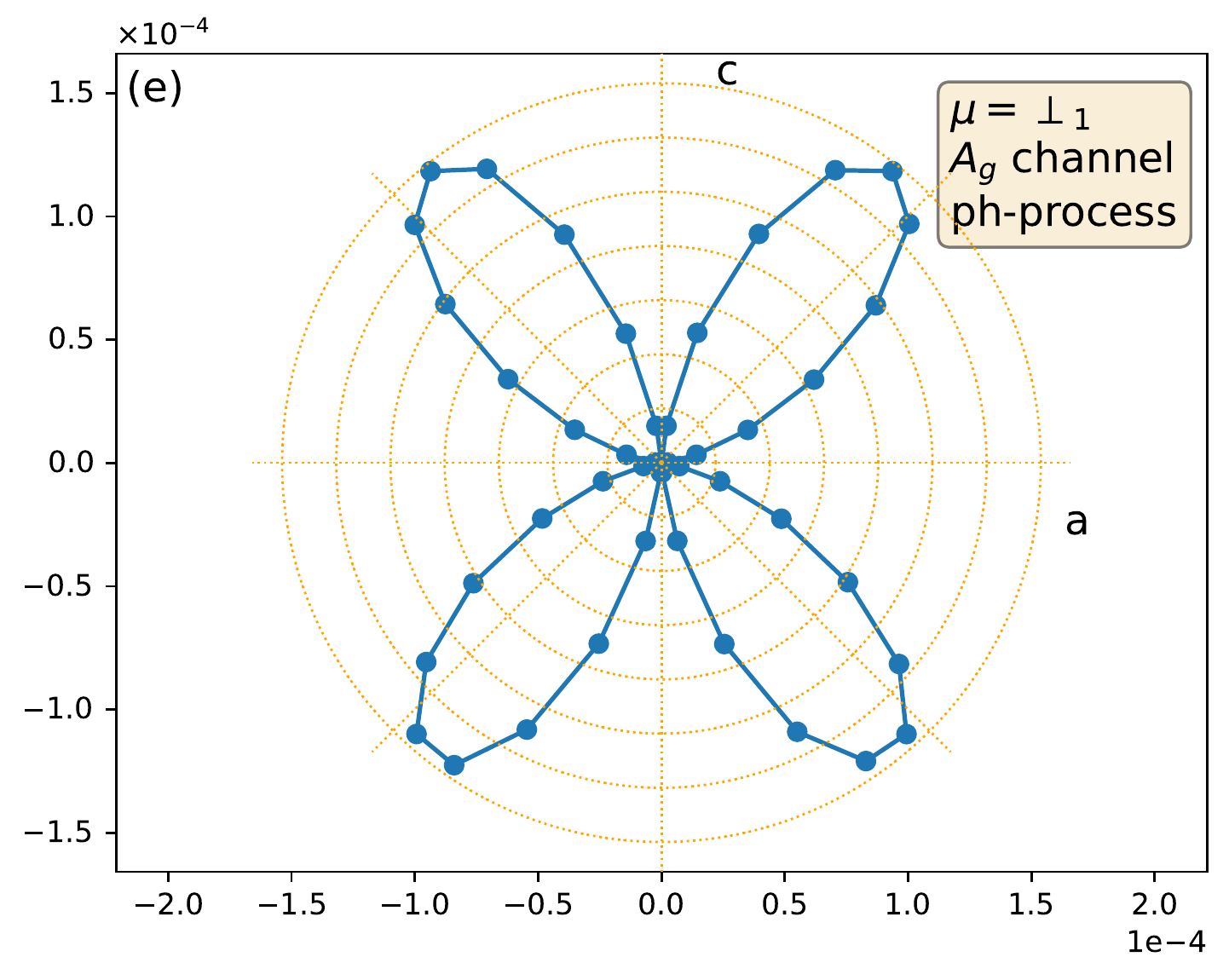}
	\caption{The angular dependence of sound attenuation coefficient 
	$\alpha_s^\mu(\theta_\vq, \phi_\vq\!=\!0^{\circ})$ in the $ac$-plane. The pp-processes contribute to  (a) the attenuation   of the longitudinal phonon ($\alpha_s^{\|}$) in $B_{2g}$ channel, (b) 
	 the attenuation of the in-plane transverse  phonon ($\alpha_s^{\perp_1}$) in the $B_{2g}$ channel, (c)  the attenuation of the out-of-plane transverse  phonon ($\alpha_s^{\perp_2}$) in the $B_{3g}$ channel. The ph-processes contribute to  (d) the attenuation of the longitudinal phonon ($\alpha_s^{\|}$)
	 and (e) of the in-plane transverse  phonon ($\alpha_s^{\perp_1}$) in the $A_g$ channel. 
	The radius represents the magnitude of $\alpha_s^\mu$ in the units of $10^{-5} \rho\delta_V$. The calculation is performed  at $T=0.02\, J$.
	}
	\label{fig: ac_pp_ph}
\end{figure*}

\subsection{Numerical results} \label{Sec:numerical}
Considering the estimations above,  we set $v_s^{\|}=3 \,J\ell$  and  $v_s^{\perp_{1,2}}\approx 1.6 \,J\ell$. 
We also take $T=0.02\, J$, which is below the flux energy gap. In the long wavelength limit, the angular dependence of the sound attenuation coefficient is scale invariant and is more experimentally relevant than the  dependence on the magnitude of the momentum  $q$.  Thus,  we fix $q=0.005 \, \ell^{-1}$ and show the polar plots of the angular dependence  of the sound attenuation (where the radius represents the magnitude of the sound attenuation coefficient).
This angular dependence is a direct reflection of the Majorana-phonon couplings (\ref{eq:symm3D}) constructed based on symmetry.

\begin{figure*}[!t]
	\centering\includegraphics[width=0.3\textwidth]{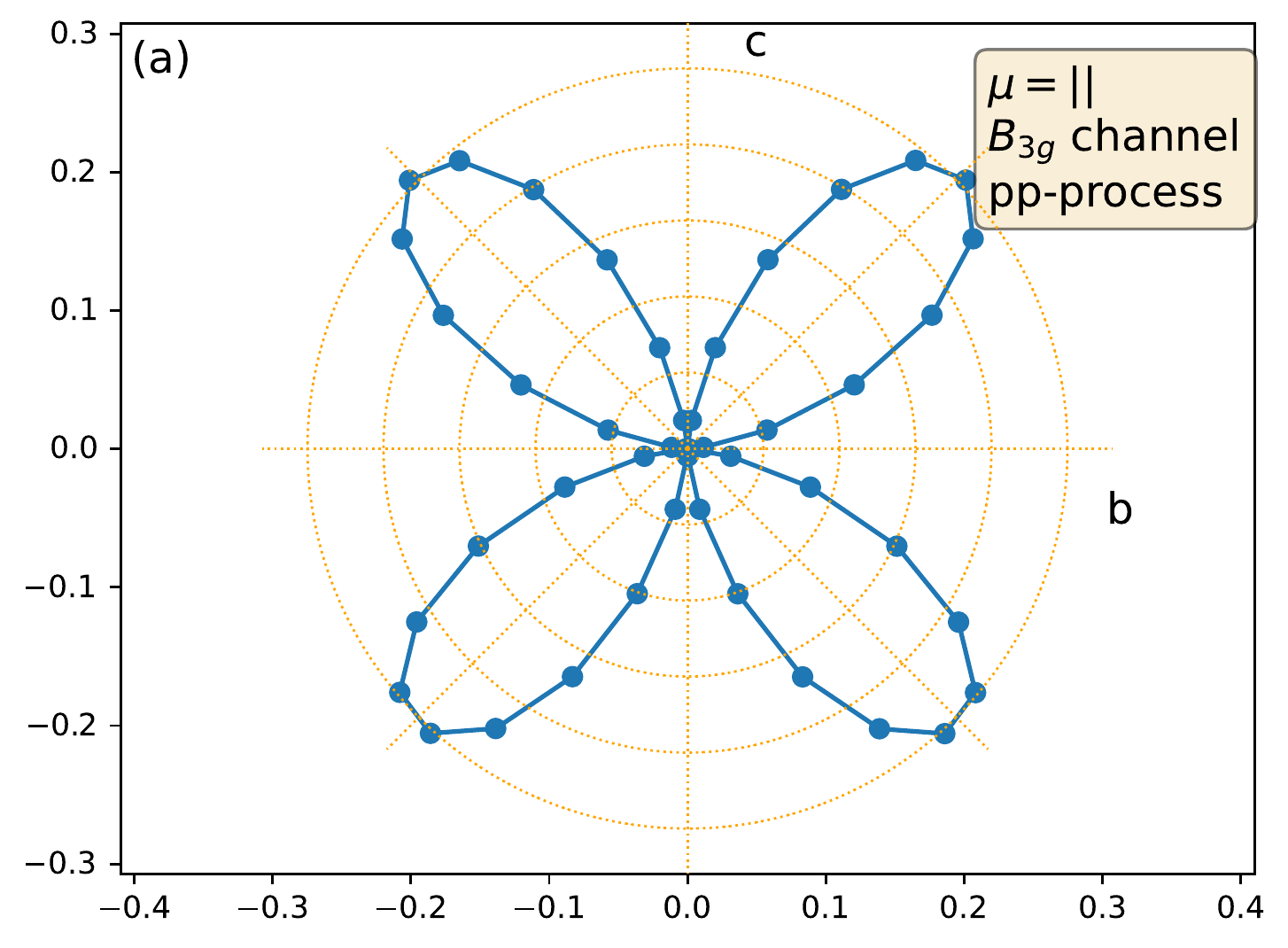}
	\includegraphics[width=0.31\textwidth]{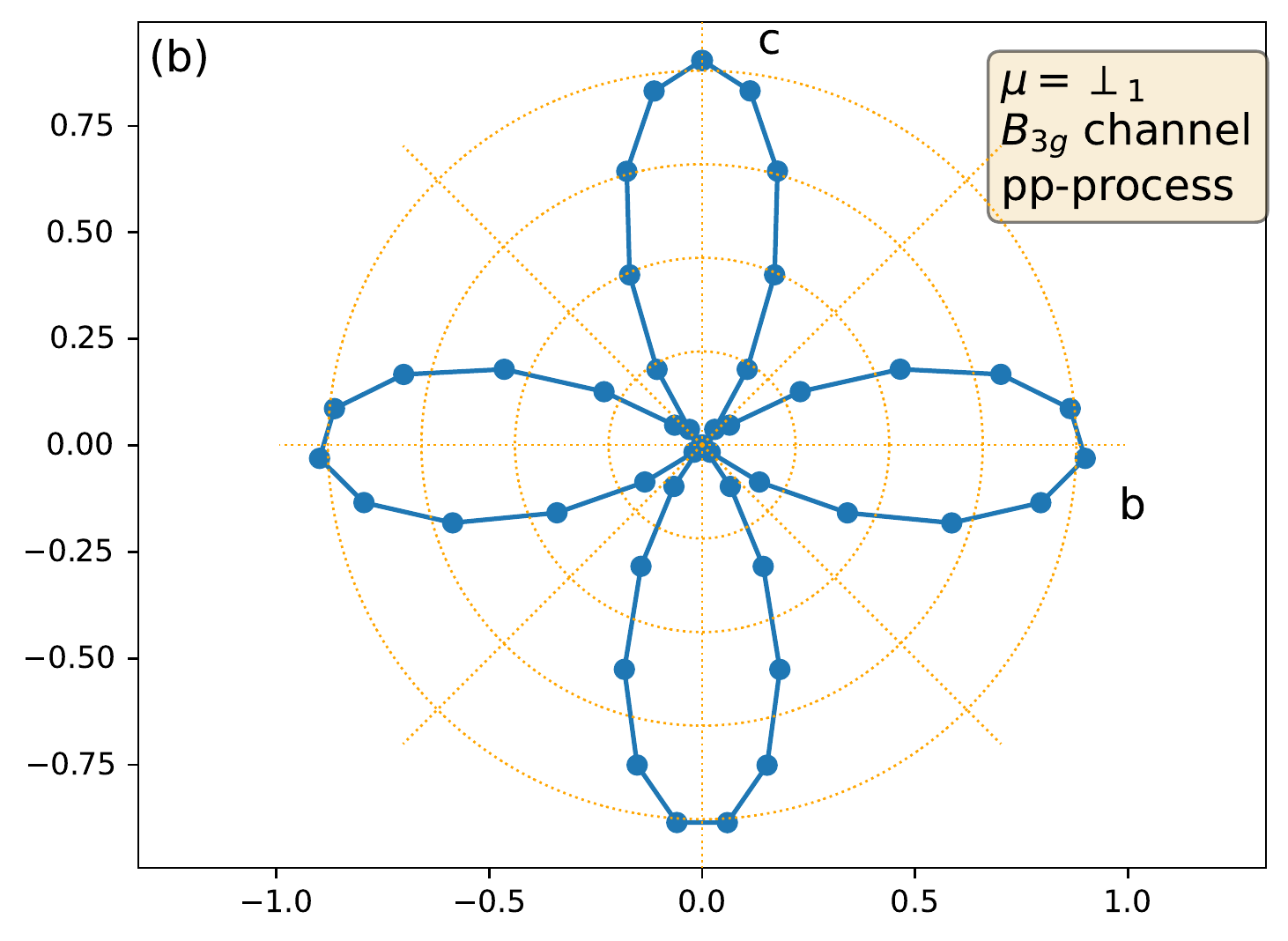}
	\includegraphics[width=0.3\textwidth] {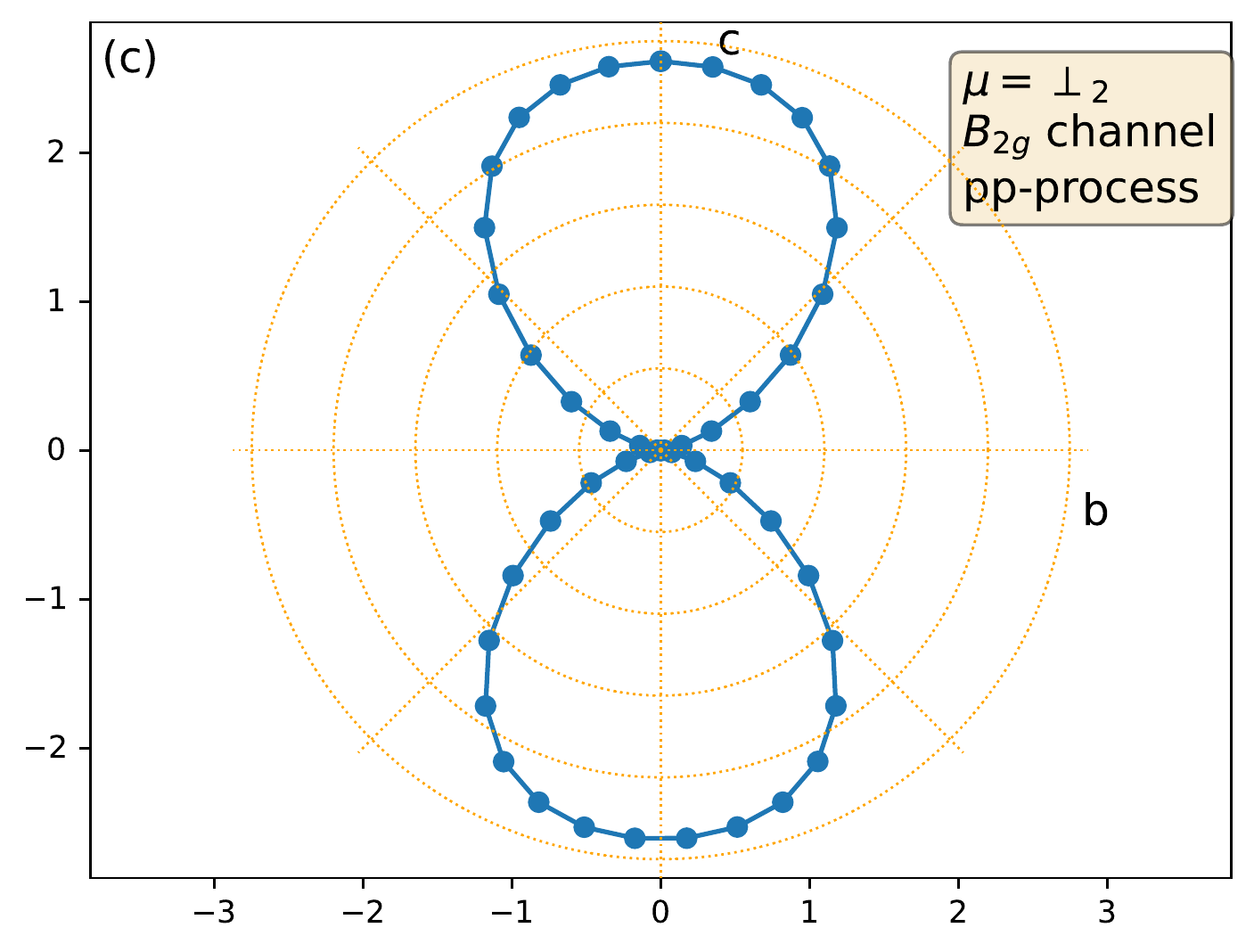}
	\includegraphics[width=0.3\textwidth]{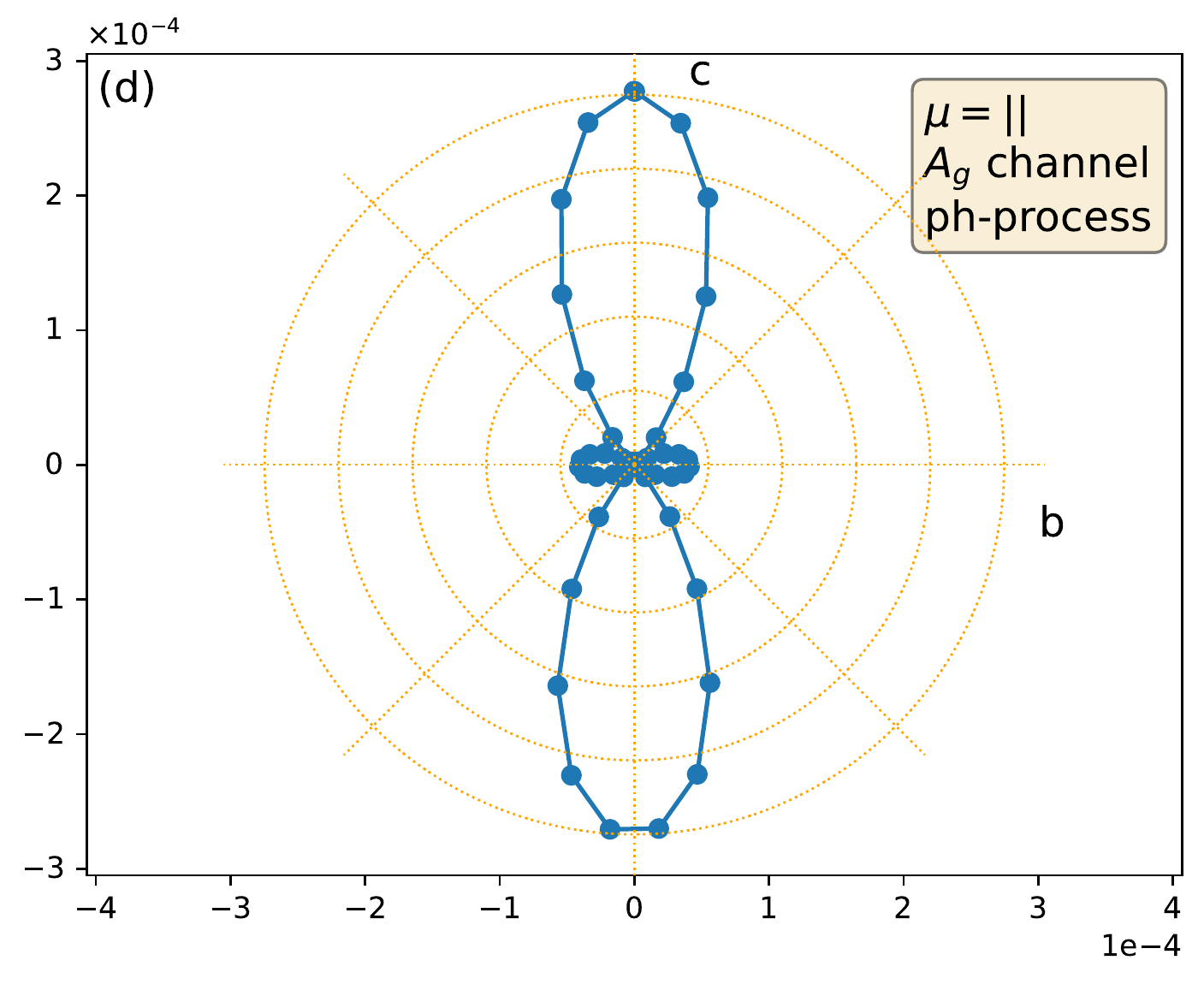}
	\includegraphics[width=0.3\textwidth]{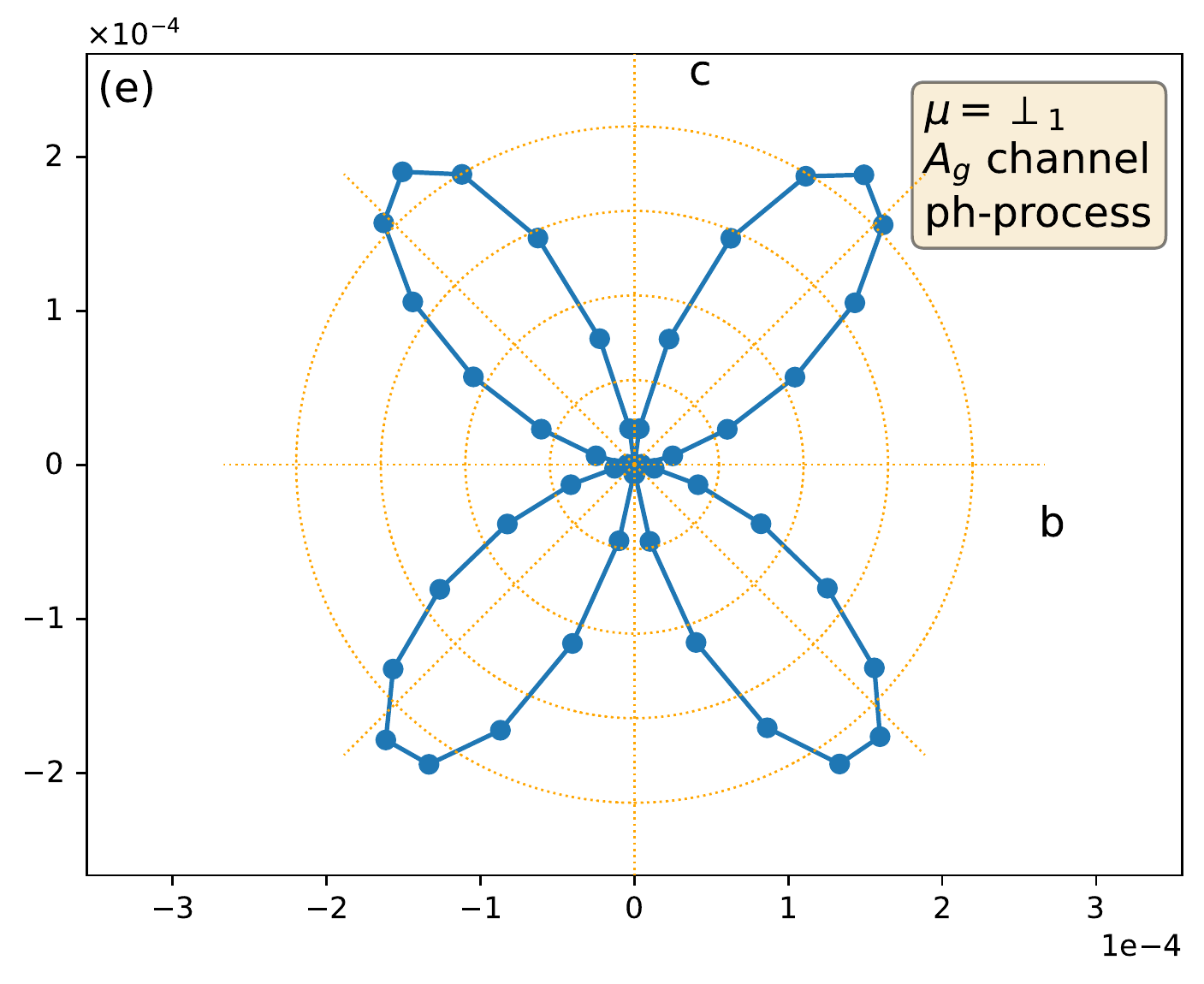}
	\caption{The angular dependence of sound attenuation coefficient 
	$\alpha_s^\mu(\theta_\vq, \phi_\vq\!=\!90^{\circ})$ in the $bc$-plane. The pp-processes contribute to  (a)   the attenuation of the longitudinal phonon ($\alpha_s^{\|}$)  in $B_{3g}$ channel, (b) the attenuation 
	of the in-plane transverse  phonon ($\alpha_s^{\perp_1}$) in the $B_{3g}$ channel, (c)  the attenuation of the out-of-plane transverse  phonon ($\alpha_s^{\perp_2}$)  in the $B_{2g}$ channel. The ph-processes contribute to (d) the attenuation of the longitudinal phonon $(\alpha_s^{\|}$) and (e) of the in-plane transverse  phonon ($\alpha_s^{\perp_1}$) in  the $A_g$ channel. 
	The radius represents the magnitude of $\alpha_s^\mu$ in the units of $10^{-5} \rho\delta_V$. The calculation is performed  at $T=0.02\, J$.
	}
	\label{fig: bc_pp_ph}
\end{figure*}

We compute  the sound attenuation coefficient in the four symmetry channels, $A_g, B_{1g}, B_{2g}, B_{3g}$,  considering separately the contributions  from  the  pp- and ph-scattering processes. In Figs.~\ref{fig: ab_pp_ph}, \ref{fig: ac_pp_ph} and \ref{fig: bc_pp_ph}, we present our results for  the sound attenuation's angular dependence patterns for the phonon modes in the three crystallographic planes, correspondingly, $ab$, $ac$ and $bc$, for three phonon's polarizations, $\|, \perp_1, \perp_2$.
The explicit expressions for the Majoarana-phonon coupling vertices in these special geometries are presented in \refapp{app: coupling-polarizations}.
These expressions show that for each of the phonon polarizations, the coupling vertex  has contributions from only two symmetry channels and another two symmetry channels give exactly zero contribution.
 Furthermore, we find that some symmetry channels have higher order (dominant) contributions in the long wavelength limit $q\to 0$. So, below we will only show the results from the leading order contributions into sound attenuation for each crystallographic plane.

{\it Phonon within the $ab$ plane}.--
 The contributions from the  pp- and ph-scattering processes for the attenuation coefficient for the phonon propagating in the $ab$ plane are shown in \reffg{fig: ab_pp_ph} (a)-(c) and (d), (e), respectively. This plane is special compared with $bc$ and $ac$ planes, because of the presence of the nodal line in the fermionic spectrum [see \reffg {fig:spectrum} (a)]  as well as the  crystallographic structure shown in \reffg{fig:HClattice}. As such, there always exist zero Fermi velocities  along the nodal line
 and small Fermi velocities in the vicinity of the nodal line. Therefore, the sound velocities along these directions are larger than the Fermi velocities, which gives rise to the non-zero  pp-processes.
In this scattering geometry, the pp-processes  contribute only in the attenuation  of the  out-of-plane transverse phonon mode in the $\perp_2$  polarization, $\alpha^{\perp_2}_s(\vq)$.  As follows from \refeq{abtr2}, $\alpha^{\perp_2}_s(\vq)$ has two contributions, one from the $B_{2g}$ channel [\reffg{fig: ab_pp_ph} (a)], describing the attenuation of lattice vibrations in the $ac$-plane, and from the $B_{3g}$ channel [\reffg{fig: ab_pp_ph} (b)], describing the attenuation of lattice vibrations in the $bc$-plane, with the former being  a bit  stronger.  The total sound attenuation of the  out-of-plane transverse phonon mode  $\alpha^{\perp_2}_s(\vq)$   shown in \reffg{fig: ab_pp_ph} (c) is the sum of  these two contributions, and its angular dependence looks like two-fold symmetric four-petal  pattern. 
 
 Since 
$v_s^{\|}\lesssim \max(v_F)$ and $v_s^{\perp_{1,2}} < \max(v_F) $, the ph-processes are also allowed.
They contribute to the  attenuation  of the longitudinal  phonon mode, $\alpha^{\|}_s(\vq)$, shown in \reffg{fig: ab_pp_ph} (d) and  of  the in-plane transverse mode, $\alpha^{\perp_1}_s(\vq)$ shown in \reffg{fig: ab_pp_ph} (e).  According to the form of Majoarana-phonon coupling vertices in this geometry given by \refeq{ablong} and \refeq{abtr1}, the attenuation of both the longitudinal and   the in-plane transverse phonons comes from the 
  dominant $A_{g}$ and subdominant $B_{1g}$ channels. However,  at   $T=0.02\, J$ both contributions are very small compared with the one from the pp-proecesses. Thus in \reffg{fig: ab_pp_ph} (d) and (e) we  only show  the angular dependence of the attenuation computed from the  $A_{g}$ contribution,  which displays a  vertical dumbbell pattern for  $\alpha^{\|}_s(\vq)$,
  and  the diagonal  four-petal pattern for $\alpha^{\perp_1}_s(\vq)$. 

Note, however, that the comparison of the magnitude of the sound attenuation coefficients between ph- and pp- processes needs to take  into consideration the temperature effect \cite{Feng2021}.
Since  the pp-processes do not require finite particle number, the low-temperature scaling behaviour of its contribution to the sound attenuation coefficient doesn't depend on temperature, i.e.\ $\alpha^{\text{pp}}_s\sim T^0$. The ph-processes require finite particle occupation, and 
its low-temperature scaling behaviour is $\alpha^{\text{ph}}_s \sim T^1$ same as in the 2D Kitaev model \cite{Ye2020}.
This is a direct result of the fact that the low-energy $\tx{DOS}(E)\sim E^1$ as shown in \reffg{fig:spectrum}(c). This low-energy scaling behaviour of $\tx{DOS}$ can also be analytically obtained by evaluating $\tx{DOS}(E) = \int_\tx{BZ} \ud^3\vk \, \delta(E - \varepsilon_{\vk})$. Then at low energy, if we expand the fermionic spectrum around the nodal line, then
\begin{align}
    \tx{DOS}(E) = \int_\tx{BZ} \ud^3 \vk \, \delta(E - v_F(k_\phi, \delta k_\theta) \, \delta k_r), \label{eq: DOS}
\end{align}
where $k_\phi$ uniquely specifies a point on the nodal line by its orientation $\phi$. Around this nodal point, on a neighboring disk locally perpendicular to the nodal line, $(\delta k_\theta, \delta k_r)$ uniquely specifies the $\vk$ point that contributes to the DOS. Then the integration \refeq{eq: DOS} is equivalent to stringing the local disks together along the nodal line. So it is easy to see that the low-energy scaling behaviour of $\tx{DOS}(E)$ is decided by the co-dimension, \ie the dimension of the BZ space minus the nodal dimension. Thus, the low-energy behaviour of $\text{DOS}(E)\sim E^1$ is the same for both 2D plane model and 3D hyperhoneycomb model, so is the low-temperature behaviour of the sound attenuation coefficient.

The low-temperature behaviours of both pp- and ph-processes distinguish themselves from the attenuation of other interaction channels, such as the channel due to phonon-phonon interactions which scales as as $\sim T^3$ in 2D and
$\sim T^5$ in 3D,  so they are promising for experimental detection at low enough temperature.

Our numerical calculation  shows that, even though the temperature dependence of attenuation from ph-process has larger power than that from pp-process, pp-process still dominates at high temperatures. The main reason is that Fermi velocities range from $0$ to $\max(v_F) = 3 J\ell$, so the sound velocities $v_s^{\|}=3 \,J\ell$ and  $v_s^{\perp_{1,2}}\approx 1.6 \,J\ell$, which  we use to describe the phonons in  $\beta$-Li$_2$IrO$_3$ compound, are still larger than a significant portion of Fermi velocities, which is consistent  with an existence of  the nearly-zero Fermi velocities along the nodal line. If we use fictitious smaller sound velocities, the contribution from the ph-processes will become larger \cite{Feng2021}.

{\it Phonon within the $ac$ or  $bc$ planes}.--
As shown in \reffg{fig: ac_pp_ph} and \reffg{fig: bc_pp_ph}, the attenuation  of the phonons propagating in the $ac$ and $bc$ planes are similar,  which is consistent with the crystallographic structure displayed in \reffg{fig:HClattice}.  Because the Fermi velocities for small deviations ${\bf k}$  from the nodal line ${\bf K}_0=(k_a,k_b,0)$ either into the $ac$ or  into the $bc$ planes are small,  at low temperatures the pp-processes dominate over the ph-processes and thus define the angular dependence of the sound attenuation.   In both geometries, the pp-processes contribute to the attenuation of the phonons with all three different polarizations angular, with similar angular patters.
  However, while for the phonon in the $ac$  plane, the strongest  attenuation is for the in-plane transverse  phonon ($\alpha_s^{\perp_1}$), for the   phonon in the $bc$  plane, the strongest  attenuation  is  for out-of-plane transverse  polarization ($\alpha_s^{\perp_2}$). In both geometries, attenuation  of phonons with out-of-plane transverse  polarization only happen through pp-processes and displays the vertical dumbbell pattern.  The four-petal angular patterns of attenuation of the  longitudinal and the in-plane transverse  phonons are rotated by 45$^\circ$ with respect to each  other.
As mentioned before, these distinct patterns directly reflect the spin-phonon couplings from different symmetry channels, probed by different phonon polarization modes.
The temperature dependence of the sound attenuation of the phonons propagating in $ac$ or $bc$ planes is similar to that in $ab$ plane.




\section{Summary}\label{sec:discussion}
In this paper, we  studied the  three-dimensional Kitaev spin-phonon model on the hyperhoneycomb lattice.
 In this model, the sound attenuation is determined by the decay of a phonon into a pair of Majorana fermions and can be  calculated from the imaginary part of the  phonon self-energy, which at the lowest order is given by  the polarization bubble.
Thus, we argued that the phonon attenuation,  measurable by the ultrasound  experiments, can serve
as an effective indirect probe of the  spin fractionalization.

In our work we considered only low temperatures below the flux disordering transition \cite{Nasu2014b},  in which only Majorana fermions contribute to the phonon self-energy. 
We showed that Majorana semimetal with nodal line band structure leaves distinct characteristic fingerprints in the temperature dependence of the phonon attenuation coefficient as a function of incident phonon momentum. First, it allows the presence of the pp-processes of the  phonon decay in  all three considered scattering geometries with the phonon propagating in one of the three crystallographic  planes. Second,  since the pp-processes of the  
 phonon decay  is allowed at all temperatures, the sound attenuation is  non zero even at zero temperature and is almost temperature independent ($\sim T^0$) at lowest temperatures.  Combining both pp-processes and ph-processes that are allowed by symmetry constraints for each scattering geometry and phonon polarization, the temperature dependence of attenuation coefficient can be schematically described by $a_T T^0 + b_T T^1$ with  $a_T> b_T$. Thus, the sound attenuation contributed from the decay into fractionalized excitations   will  be the dominant one at low enough temperatures, distinguishing itself from the contribution due to the phonon-phonon interactions, which scales as $\varpropto T^5$ in the three-dimensional system.
 We anticipate  that  the  $Z_2$ fluxes will play an important role on the phonon dynamics
  at temperatures above the flux ordering transition temperature.
  We also obtained that the sound attenuation   shows a strong  angular dependence at the leading order in  phonon momentum $q$. It is determined by the anisotropic form of the MFPh
coupling and the nodal structure of the low-energy fermionic excitations. 

Finally, we note that our study was performed for the pure
Kitaev model. Of course, real Kitaev materials feature additional weak time-reversal-invariant non-Kitaev interactions, which  give rise to other magnetic phases competing with the Kitaev  spin liquid. In particular, the  minimal spin Hamiltonian for the $\beta$-Li$_2$IrO$_3$ compound in addition   to the Kitaev coupling has contains  antiferromagnetic Heisenberg interaction  and off-diagonal $\Gamma$ exchange term~\cite{Ducatman2018}. Nevertheless, we believe that the temperature evolution of the
sound attenuation  will remain similar to the one in the pure Kitaev model as long as these perturbations do not break time reversal symmetry protecting the nodal line \cite{Hermanns2016} and
are small enough  that the material is in the proximity to the spin liquid phase.

\vspace*{0.3cm} 
\noindent{\it  Acknowledgments:} 
We thank  Rafael Fernandes,  Gabor Halasz and  Mengxing Ye for earlier  collaborations related to the topic of this study. The work of  K.F. and N.B.P. was supported by the U.S. Department of Energy, Office of Science, Basic Energy Sciences under Award No. DE-SC0018056.


\begin{widetext}
\appendix	
\section{Details of the MFPh coupling's derivation} \label{app: coupling-details}
In this appendix we present the technical details of the derivation of the  Majorana fermion-phonon (MFPh) coupling. 
  In the momentum space,  the  Majorana-phonon coupling Hamiltonian is can be written as 
\begin{eqnarray}
{ \mathcal H}^c = \sqrt{\frac{2}{N}}
\sum_{{\bf q}, {\bf k}}  ({ \mathcal H}_{{\bf q}, {\bf k}}^{A_{g}} + { \mathcal H}_{{\bf q}, {\bf k}}^{B_{1g}}
+ { \mathcal H}_{{\bf q}, {\bf k}}^{B_{2g}} +{ \mathcal H}_{{\bf q}, {\bf k}}^{B_{3g}}),
\end{eqnarray}
where the explicit expressions for the contributions from different symmetry channels are  given
by
	\begin{align}\label{Hqk}
		&{ \mathcal H}_{{\bf q},\bf{k}}^{A_{g}} =  i {\lambda_{A_g}} \bigg(4  q_c u_{{\bf q},c}
		{ \bf A}^\T_{-{\bf q}-\bf{k}}S_{\bf{k}}^{\dagger} 
		\begin{pmatrix}
			\hat{O} & -i \hat{\sigma}_3 \\
			i \hat{\sigma}_3 & \hat{O}
		\end{pmatrix} 
		S_{\bf k}
		{ \bf A}_{\bf k} + 
		(q_c u_{{\bf q},c} + 2 q_b u_{{\bf q},b}  +  
		q_a u_{{\bf q},a})  { \bf A}^\T_{{-{\bf q}-{\bf k}}} S_{\bf k}^{\dagger} \hat{Q}_{{\bf k},1}
		S_{\bf k }{\bf A}_{\bf k }\bigg),\nonumber\\\nonumber
	&	{ \mathcal H}_{{\bf q},{\bf k}}^{B_{1g}} =  \frac{i {\lambda_{B_{1g}}}  }{2}  (q_a u_{{\bf q},b} + q_b u_{{\bf q},a}) \,  { \bf A}^\T_{-{\bf q}-{\bf k}}S_{\bf k }^{\dagger} 
		\hat{Q}_{{\bf k},2}
		S_{\bf{k}}  { \bf A}_{\bf k},\\\nonumber
		&{ \mathcal H}_{{\bf q},{\bf k}}^{B_{2g}} =  
		\frac{i {\lambda_{B_{2g}}}}{2}  (q_a u_{{\bf q},c} + q_c u_{{\bf q},a}) \, { \bf A}_{-{\bf q}-{\bf k}}^\T S_{\bf k }^{\dagger}
		\hat{Q}_{{\bf k},3}
		 S_{\bf k} { \bf A}_{\bf k} , \\\nonumber
	&	{ \mathcal H}_{{\bf q},{\bf k}}^{B_{3g}} = \frac{i {\lambda_{B_{3g}}} }{2}
		 (q_b u_{{\bf q},c} + q_c u_{{\bf q},b})\,{ \bf A}^\T_{-{\bf q}-{\bf k}} S_{\bf k}^{\dagger} \hat{Q}_{{\bf k},4}
		 S_{\bf k } { \bf A}_{\bf k}.
	\end{align}
Here   $S_{\bf{k}} = {\rm diag}\{ e^{i \bf{k} \cdot {\bf r}_\alpha}\}_{\alpha = C,B,D,A}$ is the diagonal matrix in the sublattice basis,  
$\hat{O} = \begin{pmatrix}
	0 & 0  \\
	0 & 0 
\end{pmatrix} $ is the zero 2 by 2 matrix,
 $\hat{\sigma_i}$ are the  auxiliary Pauli matrices, and the explicit expressions for  $\hat{Q}_{{\bf k}}$-matrices are  given  by
\begin{align} 
&\hat{Q}_{{\bf k},1}=\left(\begin{array}{cc}
			 (1 + \cos({\bf k} \cdot {\bf a}_3)) \hat{\sigma_2} + \sin({\bf k} \cdot {\bf a}_3) \hat{\sigma_1} & \hat{O} \\
			\hat{O}&-  (\cos({\bf k} \cdot {\bf a}_1)+ \cos({\bf k} \cdot {\bf a}_2)) \hat{\sigma_2} 
			-  (\sin({\bf k} \cdot {\bf a}_1) + \sin({\bf k} \cdot {\bf a}_2)) \hat{\sigma_{1}}
		\end{array} \right),\nonumber\\
&		\hat{Q}_{{\bf k},2}=\left(\begin{array}{cc}
			 (1 + \cos({\bf k} \cdot {\bf a}_3)) \hat{\sigma_2} + \sin({\bf k} \cdot {\bf a}_3) \hat{\sigma_1} & \hat{O} \\
			\hat{O} & (\cos ({\bf k} \cdot {\bf a}_1)+ \cos({\bf k} \cdot {\bf a}_2)) \hat{\sigma_2} 
			 + (\sin({\bf k} \cdot {\bf a}_1) + \sin({\bf k} \cdot {\bf a}_2)) \hat{\sigma_{1}}		\end{array} \right),
			 			 \\\nonumber
			 &		\hat{Q}_{{\bf k},3}= \left(\begin{array}{cc}
			 (1-\cos({\bf k} \cdot {\bf a}_3)) \hat{\sigma_2} - \sin({\bf k} \cdot {\bf a}_3) \hat{\sigma_1} & \hat{O} \\
			\hat{O} &  (\cos({\bf k} \cdot {\bf a}_1) - \cos({\bf k} \cdot {\bf a}_2)) \hat{\sigma_2} 
			+ (\sin({ \bf k} \cdot {\bf a}_1) - \sin({\bf k} \cdot {\bf a}_2)) \hat{\sigma_{1}}
		\end{array} \right),\\\nonumber
		 &		\hat{Q}_{{\bf k},4}=\left(\begin{array}{cc}
			 (1-\cos({\bf k} \cdot {\bf a}_3)) \hat{\sigma_2} - \sin({\bf k} \cdot {\bf a}_3) \hat{\sigma_1} & \hat{O} \\
			\hat{O} & (\cos ({\bf k} \cdot {\bf a}_2) - \cos({\bf k} \cdot {\bf a}_1)) \hat{\sigma_2} 
			+ (\sin({\bf k} \cdot {\bf a}_2) - \sin({\bf k} \cdot {\bf a}_1)) \hat{\sigma_{1}}
		\end{array} \right).
\end{align}
Next we rewrite ${\mathcal H}^c$ in terms of the transverse and longitudinal eigenmodes as in  \refeq{Hcpolarizations}, where  corresponding  MFPh coupling vertices  are  given by
	\begin{align}
	\hat{\lambda}_{{\bf q},\bf{k}}^{\parallel} &= 
		 i \lambda_{A_g} \bigg({4 q_c R_{31}  \begin{pmatrix}
			\hat{O} & -i \hat{\sigma}_3 \\
			i \hat{\sigma}_3 & \hat{O}
		\end{pmatrix}   
		+ (q_c R_{31} + 2 q_b R_{21} + q_a R_{11}) \,  \hat{Q}_{{\bf k},1}   }  \bigg)
		\label{generallong}\\\nonumber
		&
		+ \frac{i \lambda_{B_{1g}}}{2} (q_a R_{21} + q_b R_{11})\, \hat{Q}_{{\bf k},2}+  \frac{i \lambda_{B_{2g}}}{2} (q_a R_{31} + q_c R_{11}) \, \hat{Q}_{{\bf k},3} +
		\frac{i \lambda_{B_{3g}}}{2} ( q_b R_{31} + q_c R_{21})  \, \hat{Q}_{{\bf k},4},\\
		\label{generaltr1}
	\hat{\lambda}_{{\bf q},\bf{k}}^{\perp_1}&=
		i \lambda_{A_g}\bigg( { 4 q_c R_{32} \begin{pmatrix}
				\hat{O} & -i \hat{\sigma}_3 \\
				i \hat{\sigma}_3 & \hat{O}
			\end{pmatrix}   
			+ (q_c R_{32} + 2 q_b R_{22} + q_a R_{12}) \, \hat{Q}_{{\bf k},1}  } \bigg)\\\nonumber
		&+ \frac{i \lambda_{B_{1g}}}{2} (q_a R_{22} + q_b R_{12}) \, \hat{Q}_{{\bf k},2}+ \frac{i \lambda_{B_{2g}}}{2} ( q_a R_{32} + q_c R_{12} )  \, \hat{Q}_{{\bf k},3} +
		\frac{i \lambda_{B_{3g}}}{2} ( q_b R_{32} + q_c R_{22} )  \, \hat{Q}_{{\bf k},4},	\\
		\label{generaltr2}
	\hat{\lambda}_{{\bf q},\bf{k}}^{\perp_2} &=
		i \lambda_{A_g} \bigg({4 q_c R_{33} \begin{pmatrix}
				\hat{O} & -i \hat{\sigma}_3 \\
				i \hat{\sigma}_3 & \hat{O}
			\end{pmatrix}   
			+ (q_c R_{33} + 2 q_b R_{23} + q_a R_{13}) \, \hat{Q}_{{\bf k},1}  } \bigg )\\\nonumber
		&+ \frac{i \lambda_{B_{1g}}}{2} (q_a R_{23} + q_b R_{13}) \, \hat{Q}_{{\bf k},2}+ \frac{i \lambda_{B_{2g}}}{2} (q_a R_{33} + q_c R_{13} )  \, \hat{Q}_{{\bf k},3} +
		\frac{i \lambda_{B_{3g}}}{2} ( q_b R_{33} + q_c R_{23}) \, \hat{Q}_{{\bf k},4}.
	\end{align}
Note also  that since we are using the long wavelength limit  for the phonons, we only
 kept the  leading in $q$ terms in all the expressions.
 
\section{MFPh couplings  in  various polarizations} \label{app: coupling-polarizations}  

For phonon in  the $ab$ plane, the rotation matrix is given by
$
\hat{R} = \left[
\begin{array}{ccc}
    \cos\phi_\vq & -\sin\phi_\vq        & 0 \\
    \sin\phi_\vq  & \cos\phi_\vq                  & 0 \\
    0         & 0 & 1
\end{array}\right],
$
 which simplifies the general expressions for the MFPh coupling vertices  to 
\begin{align}
\hat{\lambda}_{\vq, \vk}^{\|}=
&
i \lambda_{A_{g}}q_{a} R_{11} \hat{Q}_{\vk, 1}+\frac{i \lambda_{B_{1 g}}}{2}\left(q_{a} R_{21}+q_{b} R_{11}\right) \hat{Q}_{\vk, 2}, 
\label{ablong}\\
\hat{\lambda}_{\vq, \vk}^{\perp_1}=
&
i \lambda_{A_{g}}\left(2 q_{b} R_{22}+q_{a} R_{12}\right) \hat{Q}_{\vk, 1} +\frac{i \lambda_{B_{1 g}}}{2}\left(q_{a} R_{22}+q_{b} R_{12}\right) \hat{Q}_{\vk, 2} 
\label{abtr1}\\
\hat{\lambda}_{\vq, \vk}^{\perp_2}=
&
\frac{i \lambda_{B_{2 g}}}{2}q_{a} R_{33} \hat{Q}_{\vk, 3}+\frac{i \lambda_{B_{3 g}}}{2}q_{b} R_{33} \hat{Q}_{\vk, 4}.\label{abtr2}
\end{align}

Similarly, for the phonon in the $ac$ plane,  the rotation matrix is given by
$\hat{R}  = \left[
\begin{array}{ccc}
    \sin\theta_\vq & \cos\theta_\vq          & 0 \\
    0 & 0                   & 1 \\
    \cos\theta_\vq         & -\sin\theta_\vq  & 0
\end{array}\right]$, so the  MFPh coupling vertices are given by
\begin{align}
\hat{\lambda}_{\vq, \vk}^{\|}=& i \lambda_{A_{g}}\left(4 q_{c} R_{31}\left(\begin{array}{cc}
    \hat{O} & -i \hat{\sigma}_{3} \\
    i \hat{\sigma}_{3} & \hat{O}
    \end{array}\right)+\left(q_{c} R_{31}+q_{a} R_{11}\right) \hat{Q}_{\vk, 1}\right)
    +\frac{i \lambda_{B_{2 g}}}{2}\left(q_{a} R_{31}+q_{c} R_{11}\right) \hat{Q}_{\vk, 3}, 
\\
\hat{\lambda}_{\vq, \vk}^{\perp_1}=& i \lambda_{A_{g}}\left(4 q_{c} R_{32}\left(\begin{array}{cc}
    \hat{O} & -i \hat{\sigma}_{3}\\
    i \hat{\sigma}_{3} & \hat{O}
    \end{array}\right)+\left(q_{c} R_{32}+q_{a} R_{12}\right) \hat{Q}_{\vk, 1}\right) 
    +\frac{i \lambda_{B_{2 g}}}{2}\left(q_{a} R_{32}+q_{c} R_{12}\right) \hat{Q}_{\vk, 3}, 
\\
\hat{\lambda}_{\vq, \vk}^{\perp_2}= &\frac{i \lambda_{B_{1 g}}}{2}q_{a} R_{23} \hat{Q}_{\vk, 2}+
\frac{i \lambda_{B_{3 g}}}{2}q_{c} R_{23} \hat{Q}_{\vk, 4}.
\end{align}

For the phonon in the $bc$ plane, the rotation matrix is
$\hat{R}  = \left[
\begin{array}{ccc}
    0 & 0        & 1 \\
    \sin\theta_\vq & \cos\theta_\vq    & 0 \\
    \cos\theta_\vq        & -\sin\theta_\vq & 0
\end{array}\right]$
, and the  MFPh coupling vertices are given by
\begin{align}
\hat{\lambda}_{\vq, \vk}^{\|}=& i \lambda_{A_{g}}\left(4 q_{c} R_{31}\left(\begin{array}{cc}
    \hat{O} & -i \hat{\sigma}_{3} \\
    i \hat{\sigma}_{3} & \hat{O}
    \end{array}\right)+\left(q_{c} R_{31}+2 q_{b} R_{21}\right) \hat{Q}_{\vk, 1}\right)
    +\frac{i \lambda_{B_{3 g}}}{2}\left(q_{b} R_{31}+q_{c} R_{21}\right) \hat{Q}_{\vk, 4}, 
\\
\hat{\lambda}_{\vq, \vk}^{\perp_1}=& i \lambda_{A_{g}}\left(4 q_{c} R_{32}\left(\begin{array}{cc}
    \hat{O} & -i \hat{\sigma}_{3}\\
    i \hat{\sigma}_{3} & \hat{O}
    \end{array}\right)+\left(q_{c} R_{32}+2 q_{b} R_{22}\right) \hat{Q}_{\vk, 1}\right) 
    +\frac{i \lambda_{B_{3 g}}}{2}\left(q_{b} R_{32}+q_{c} R_{22}\right) \hat{Q}_{\vk, 4}, 
\\
\hat{\lambda}_{\vq, \vk}^{\perp_2}=
    &\frac{i \lambda_{B_{1 g}}}{2}q_{b} R_{13} \hat{Q}_{\vk, 2}+\frac{i \lambda_{B_{2 g}}}{2}q_{c} R_{13} \hat{Q}_{\vk, 3}.
\end{align}

So in each plane, only two of the fours symmetry channels are active. And as  shown in the numerical calculations presented in the main text, in the long wavelength limit, one of the two channels dominates over the other. Similar situation was observed in the analysis of the 2D spin-phonon Kitaev model \cite{Ye2020}.

\section{Explicit expressions for the dynamical factors in (20)}\label{app:dynamicalfactor}

 The dynamic factors in Eq.~(20)  are evaluated as follows:
\begin{align}\label{eq:PPPP}
    P_{\vk, 11} &= T \sum_{i \omega_{m}} \frac{1}{\left(i \Omega_{n}+i \omega_{m}\right)-\varepsilon_{\vk}}\frac{1}{i \omega_{m}-\varepsilon_{\vk + \vq}} =\frac{n_{F}\left(\varepsilon_{\vk}\right)-n_{F}\left(\varepsilon_{\vk + \vq}\right)}{i \Omega_{n}-\varepsilon_{\vk}+{\varepsilon_{\vk + \vq}}},\nonumber
    \\
    P_{\vk, 22} &=T\sum_{i \omega_{m}}\frac{1}{\left(i\Omega_{n}+i \omega_{m}\right) +\varepsilon_{\vk}}\frac{1}{i \omega_{m}+\varepsilon_{\vk + \vq}} = \frac{n_{F}\left(-\varepsilon_{\vk }\right)-n_{F}\left(-\varepsilon_{\vk+ \vq}\right)}{i \Omega_{n}+\varepsilon_{\vk}-\varepsilon_{\vk + \vq}} ,
    \\
    P_{\vk, 21}  &=T\sum_{i \omega_{m}}\frac{1}{\left(i \Omega_{n}+i \omega_{m}\right)+\varepsilon_{\vk}} \frac{1}{i \omega_{m}-\varepsilon_{\vk + \vq}}  = \frac{n_{F}\left(-\varepsilon_{\vk }\right)-n_{F}\left(\varepsilon_{\vk+ \vq}\right)}{i \Omega_{n}+\varepsilon_{\vk }+\varepsilon_{\vk+ \vq}} ,
    \\
    P_{\vk, 12} &=T\sum_{i \omega_{m}}\frac{1}{\left(i \Omega_{n}+i \omega_{m}\right)-\varepsilon_{\vk}}\frac{1}{i \omega_{m}+\varepsilon_{\vk + \vq}} =\frac{n_{F}\left(\varepsilon_{\vk}\right)-n_{F}\left(-\varepsilon_{\vk + \vq}\right)}{i \Omega_{n}-\varepsilon_{\vk }-\varepsilon_{\vk+ \vq}}, \nn
\end{align}

\section{Vegas+ Monte Carlo integration}

In this paper, we applied  an efficient Monte Carlo algorithm for multidimensional integration Vegas+ \cite{lepage2021adaptive, dehesa2011quantum} to evaluate the phase space integration in the polarization bubble \refeq{eq: bubble}. In this section, we will briefly discuss the technical aspect of this algorithm.


Vegas+ is an adaptive stratified sampling algorithm, which is very effective for the integrands with multiple peaks or diagonal nodal (significant) structures. In general, an importance sampling  (as in  the original Vegas algorithm) is a basic variance reduction technique in Monte Carlo integration, where the probability space is transformed, such that the sampling is concentrated on the important region of the integrand. 
For example, suppose we need to compute a 1D integral
\begin{align}
    I = \int_a^{b} f(x) \ud x.
\end{align}
Different from directly sampling $x \in [a, b]$, as is done in a standard Monte Carlo technique, importance sampling  introduces a measurable map from $y$ to $x$, $x = G^{-1}(y)$, where $y\in [0,1]$. Then, the integration is equivalently written as:
\begin{align}
    I = \int_{0}^{1} f(x(y)) \frac{\ud x}{\ud y} \ud y,
\end{align}
and, instead of uniformly sampling $x\in [a, b]$, one uniformly samples $y \in [0, 1]$. The result of this probability space transformation is such that the distribution of $x$ is described by function $g(x) = G'(x)$ (known from inverse transform sampling). If $g(x)$ is well designed to be of similar shape to $f(x)$, \ie $g(x)$ is large where $f(x)$ is large, then the $x$ samples will be concentrated in the important region of $f(x)$. 

What the Vegas algorithm \cite{lepage1978new} does is to numerically obtain the map $G^{-1}: y \to x$, which gives the probability distribution function $g(x)$, in the following adaptive way.
First the $x$ integration space is partitioned into $N_p$ intervals, and $\Delta x_i$ is the length of each interval (not necessarily uniform).  Then the functional form is chosen such that  $x$ monotonically increases with $y$, and within each partition of $x$, the increase is linear with a rate (Jacobian) $J_i$, \ie $\Delta x_i = J_i \Delta y_i$, (again not necessarily uniform).
So the measurable map $G^{-1}: y \to x$ is specified by the set of variables $\{\Delta x_i, J_i\}$, which are under the constraints $\sum_{i=1}^{N_p} \Delta x_i = b-a$, $\sum_{i=1}^{N_p} \Delta y_i = 1$. The objective of designing the distribution function $g(x)$ is to minimize the variation of the integrand (seen as a function of random variable $y$):
\begin{align}
    \sigma_I^2  = \mathbb{V}\tx{ar}_{y\in[0,1]}
    \left[f(x(y))\frac{\ud x}{\ud y}\right] 
               = \int_0^1 \left[ f(x(y))\frac{\ud x}{\ud y}\right]^2\ud y - I^2
               = \sum_i J_i \int_{x_i}^{x_{i} + \Delta x_i}  f(x)^2 \ud x - I^2,
\end{align}
where $x_i$ is the left end of each interval partitioned from $x \in [a, b]$. So now designing the map $G^{-1}: y \to x$ becomes a constrained optimization problem 
\begin{align}
    &\min_{\{\Delta x_i, J_i\}} \sigma_I^2 \left( \{\Delta x_i, J_i\} \right).
\end{align} 
From here, it is easy to get the necessary optimal condition \cite{lepage2021adaptive}:
\begin{align}
    \frac{1}{\Delta x_i} \int_{x_i}^{x_i+\Delta x_i} J_i^2 f(x)^2 \ud x = \tx{constant},
\end{align}
i.e. the optimal partition grid $\{x_i\}$ is such that the average of $J_i^2 f(x)^2$ over each interval $\Delta x_i$ is uniform across the partitions. Without loss of generality, we can introduce uniform grid (partition) in $y$ space, \ie $\Delta y_i = 1/N_p$. Then, $J_i=\frac{\Delta x_i}{\Delta y_i} = \Delta x_i \cdot N_p$. Then, the objective becomes finding the grid in $x$ space, such that the average of $\Delta x_i^2 f(x)^2 $ over $\Delta x_i$ is uniform, the result of which leads to importance sampling. 

The uniform $\Delta x_i^2 f(x)^2 $ is achieved by an adaptive numerical algorithm, which can be intuitively understood as follows.
First, the average $w_i = \lefta \Delta x_i^2 f(x)^2\righta_{\Delta x_i}$ on $\Delta x_i$ is defined to be the weight of the $i$-th partition.
 We also define the center weight of all partitions to be $c = \onefrac{N_p} \sum_i w_i$. Then the uniform weight $\{w_i\}$ condition is equivalent to requiring 
 $\sum_i \left|w_i - c\right|^2$  to be minimized. In other words,  we have the following optimization problem:
\begin{align}
    L(\{x_i\}) = \min_{\{x_i\}}\min_{c}\sum_i\left|w_i - c\right|^2
\end{align}
We can easily verify that uniform $\{w_i\}$ is indeed the saddle point solution, i.e., if $\{w_i\}$ is uniform, then $L=0$. This problem is solved by an alternating optimization algorithm, which alternatively updates $\{x_i\}$ and $c$ in an adaptive procedure \cite{lepage2021adaptive}. The optimal solution yields a grid of $x$, which is the most dense in the importance region of the integrand. Thus Vegas is considered an adaptive importance sampling method.



Next, we introduce vegas+, the enhanced version of vegas with stratified sampling. In the above algorithm, we have obtained the the uniform grid of $\{y_i=1/N_p\}$, so it is natural to stratify the sampling according this partition.
To obtain the optimal number of samples allocated to each stratum $\{n_i\}$, we can minimize the Monte Carlo standard deviation $\sigma_\tx{MC}^2 = \sum_{i=1}^{N_p} \frac{\sigma^2_i(fJ_i)}{n_i}$ with the constraint $\sum_i n_i = N_{total}$, where 
$\sigma_i(fJ_i)$ is the variance of $f(x)J_i$ in the $i$-th partition. This gives that the optimal stratification is $n_i \propto \sigma_i(fJ_i)$. The same optimization method was also applied in the stratified Monte Carlo simulations in the 2D Kitaev QSL ~\cite{Feng2021, feng2022phonon}, except that there $\sigma_\tx{MC}^2 = \sum_i \frac{p_i^2\sigma_i^2}{n_i}$, where $p_i$ is the normalized probability of the $i$-th partition, and the optimal stratification is $n_i \propto p_i \sigma_i \approx p_i$. 

Finally,  in this paper the integration was done in 3D ${\bf k}$-space with the important region centered around the nodal line. 
The Vegas algorithm was used to  make sure that the samples are concentrated near the 2D plane. But within that 2D plane, partition grid is basically uniform. At this point, the adaptive stratified sampling  of Vegas+  was used to assure that the dominant contribution comes from
the samples only in the important hypercubes near the nodal line.

\end{widetext}

\bibliography{References}

\begin{thebibliography}{64}
\bibitem [{Tsi()}]{Tsirlin}%
  \BibitemOpen
  \href@noop {} {}\bibinfo {note} {From private communications with Alexander
  Tsirlin, the non-zero elastic modulus tensor coefficients in
  $\beta$-LiIr$_2$O$_3$ (in kbar) are $C_{aaaa}=2848.9530$,
  $C_{bbbb}=2636.5278$, $C_{cccc}=2658.3422$, $C_{aabb}=881.9877$,
  $C_{aacc}=1273.9659$, $C_{bbcc}=1177.4125$, $C_{abab}=814.5967$,
  $C_{acac}=903.3235$, $C_{bcbc}=704.7004$.}\BibitemShut {Stop}%
\end{thebibliography}
		
\end{document}